 \documentclass[reprint]{JASA} 

\usepackage{tipa} 

\emergencystretch=\maxdimen
\begin{document}


\title[~~~]{Anisotropic minimum dissipation subgrid-scale model in hybrid aeroacoustic simulations of human phonation}


\author{Martin Lasota}
\email{Martin.Lasota@TUL.cz}
\affiliation{Institute of New Technologies and Applied Informatics, Technical University of Liberec, Studentská~2, 460 01 Liberec, Czechia}

\author{Petr \v{S}idlof}
\altaffiliation{Also at: Institute of Thermomechanics, Czech Academy of Sciences, Dolejškova 5, 182 00 Prague, Czechia}
\affiliation{Institute of New Technologies and Applied Informatics, Technical University of Liberec, Studentská~2, 460 01 Liberec, Czechia}

\author{Paul Maurerlehner}
\author{Manfred Kaltenbacher}
\author{Stefan Schoder}
\affiliation{Institute of Fundamentals and Theory in Electrical Engineering, Graz University of Technology, Inffeldgasse 18, 8010 Graz, Austria}








\begin{abstract}

This \replaced{paper}{article} deals with large-eddy simulations\deleted{(LES)} of 3D incompressible laryngeal flow followed by acoustic simulations of \added{human} phonation of five cardinal \added{english} vowels /u, i, \textipa{A}, o, æ/. \deleted{That aims to describe the correlation between turbulent flow simulations with a conventional (One-equation, WALE) or anisotropic minimum dissipation (AMD) subgrid-scale (SGS) turbulence model and its impact on the aeroacoustic spectrum in human phonation.}\added{The flow and aeroacoustic simulations were performed in OpenFOAM and in-house code openCFS, respectively.} Given the large variety of scales in the flow and acoustics, the simulation is separated into two steps: (1) computing the flow in the larynx using the finite volume method on a fine \added{2.2M} grid followed by (2) computing the sound sources separately and wave propagation to the radiation zone around the mouth using the finite element method on a coarse \added{33k} acoustic grid.\deleted{The first application of AMD in biological flow is included. For all simulated vowels, the AMD model predicted stronger second formants up to 2 kHz than the widely used wall-adapting local eddy-viscosity model. Based on our findings, the AMD model's anisotropic behavior proved advantageous in dynamic mesh applications, such as the moving grid within the glottis.} \added{The numerical results showed that the anisotropic minimum dissipation model, which is not well known since it is not available in common CFD software, predicted stronger sound pressure levels at higher harmonics and especially at first two formants than the wall-adapting local eddy-viscosity model. We implemented the model as a new open library in OpenFOAM and deployed the model on turbulent flow in the larynx with positive impact on the quality of simulated vowels. Numerical simulations are in very good agreement with positions of formants from measurements.}

\end{abstract}


\maketitle

\section{Introduction}

Aeroacoustic simulations undoubtedly have strong potential to be applied in clinical diagnostics, treatment control, and supporting medical education. Regarding this application, computer analysis can provide highly resolved 3D data of the flow and acoustic field for further studies. Such highly resolved 3D data are infeasible by in-vivo, excised larynx, or synthetic vocal fold measurements \citep{Doellinger-CB2011,Kniesburges-CB2011}. 

Although numerical tools are available, challenges exist. Realistic 3D turbulent flow simulations are computationally expensive since the time-consuming accurate modeling of supraglottal turbulence is essential for flow-induced sound generation \citep{Mattheus-JASA2011}. Derived from the fact that turbulence must be resolved, conventional turbulence modeling approaches \added{unsteady Reynolds-averaged Navier-Stokes} (uRANS) equations are inadequate for aeroacoustics because they do not provide the instantaneous state of flow quantities. Therefore, compared to a direct numerical simulation (DNS), \added{large-eddy simulation} (LES) is a beneficial balance between the resolution of turbulent structures and the accurate prediction of turbulent sound generation mechanism. One of the first studies employing LES to study laryngeal aeroacoustics was the work of \citet{Suh-JASA2007}, who combined a compressible LES and acoustic analogy published by \citet{Ffowcs-PTRSL1969} (FW-H) in a static model of the human glottis for human voice signal predictions. \citet{Mihaescu-JASA2010} employed the LES capability to study the laryngeal airflow during phonation and inspiration. \citet{Schwarze-CF2011} explored an implicit LES, where the intrinsic dissipation of the numerical method was assumed to act as a subgrid-scale model. During recent years, simulations have advanced considerably \citep{Bodaghi-JBE2021, Tokuda-JASA2017, Yoshinaga-JSV2017} towards a state where full-scale aeroacoustic simulations on realistic CT- or MRI-based geometries are possible. It can be anticipated that these simulations could be used for subject-specific pre-surgical predictions of vocal fold oscillations \citep{Avhad-F2022}. These phonation simulations are useful and help to improve the voice quality for subjects suffering from various vocal fold dysfunctions \citep{Falk-FP2021,Sadeghi-JASA2019,Sadeghi-APPSCI2019}, or evaluate potential effects on voice production affected by an implant insertion in medialization laryngoplasty \citep{Zhang-PO2020}.

In prospective clinical applications, the reliability of the LES is a key factor. Hence further studies on the accuracy of new subgrid-scale models are important. Regarding our previous LES study \citep{Lasota-APPSCI2021}, the relatively new subgrid-scale anisotropic minimum dissipation (AMD) was implemented in OpenFOAM\footnote{https://github.com/OpenFOAM} and first results were analyzed. The reasons for introducing the given model are based on the major features of the minimum dissipation models, which are constructed to confine time derivative of subgrid-scale energy, and satisfy the need to dissipate the energy of subgrid-scales. In addition, the AMD model is also derived for anisotropic meshes, which can circumvent an overuse of corrections to keep stability by numerical schemes when the fast abducting and adducting of the glottis occurs. That means AMD does not require \citep{Rozema-PF2015} approximating the \replaced{filter}{width of the LES filter $\Delta=(\Delta_x \Delta_y \Delta_z)^{1/3}$}, which is usually determined using an explicit box filter of width the mesh size. The study published by Rozema confirmed high consistency between results obtained by AMD and the DNS of turbulent channel flow, or also between AMD and the DNS of temporal mixing layer, both done on anisotropic grids. These mentioned flow characteristics are comparable to the flow structures arising inside the larynx. \replaced{Another feature is hidden in the fact that the AMD model is consistent with65
the exact subgrid-scale stress tensor $\tau_{ij}$, which means switching off the AMD model in cases of non-turbulent flows}{Another positive feature is hidden in the fact that the AMD model computes exactly same values which are obtained by computing the exact subgrid-scale stress tensor $\tau_{ij}$, and thus the AMD model allows very practically switching off in cases of non-turbulent flows.} \citet{Abkar-2017BLM} performed the computation of atmospheric boundary layer flows in the thermally stratified atmosphere with a slight modification of AMD by adding a contribution of buoyant forces, and they pointed out a good agreement with well-established empirical correlations, theoretical predictions, and field observations. In particular, they referred that the results are weakly dependent on the grid resolution, indicating the robustness of the proposed model.

In this work, the AMD model is investigated for biological laryngeal flows based on the above-mentioned advantages. The flow field prediction capabilities are assessed and compared to state-of-the-art LES using a conventional One-Equation (OE) and Wall-Adapting Local Eddy-viscosity (WALE) subgrid-scale turbulence model. Furthermore, the AMD model's impact on the aeroacoustic sound generation and the produced voice signal is evaluated for the first time.

This paper is organized into two major sections, describing the computational fluid dynamic (CFD) and \added{computational} aeroacoustic (CAA) models. Section \ref{Chap:CFD} describes the LES framework, three different SGS models, geometry \& \replaced{BCs}{boundary conditions}, CFD mesh \& discretization, numerical solution, and CFD results of the laryngeal turbulent flow. Section \ref{Chap:CAA} keeps the same structure, i.e. the CAA model, geometry \& \replaced{BCs}{boundary conditions}, acoustic mesh \& discretization, numerical solution, and CAA results of the human phonation divided into the visualization of sound sources and acoustic wave propagation (in the time and frequency domain).

\section{CFD model} \label{Chap:CFD}

 \subsection{Mathematical model} \label{Chap:CFD:Model}
 
 Large-eddy simulation (LES) is a mathematical concept for modeling turbulent flows, which deals with flow structures carrying most kinetic energy $k$, i.e., spatially organized large scales. These consist of two main categories: coherent structures and coherent vortices of recognizable shape \citep{Lesieur-book2005}. In the numerical implementation, the characteristic length $\Delta$, defining a cutoff between resolved large scales and modeled subgrid scales, is usually given by the mesh grid spacing \citep{Versteeg-book2016}. 

In the LES concept, any flow variable $f(\mathbf{x},t)$, where $\mathbf{x} = (x_1,x_2,x_3)$ is the spatial coordinate and $t$ time, may be decomposed as

\begin{equation}
f(\mathbf{x},t)=\overline{f}(\mathbf{x},t)+f'(\mathbf{x},t) ,
\end{equation}

\noindent where $\overline{f}(\mathbf{x},t)= G_{f}(\mathbf{x})*f(\mathbf{x},t)=\int G_{f}(\mathbf{r},\mathbf{x},\Delta) f(\mathbf{x}-\mathbf{r},t) {\rm d}\mathbf{r}$ is the large-scale component, obtained by spatial filtering, and $f'(x,t)$ is the small subgrid-scale contribution. Filtered variables for LES are functions of time and space, unlike the Reynolds averaged variables, hence in LES: $\overline{\overline{f}} \neq \overline{f}$, $\overline{f'} \neq 0$. The convolution introduced above contains a filter function $G_{f}$ separating spatial scales. The used filter in this study is the top-hat filter, which is a common choice in low-order finite volume methods:

\begin{equation}
    G_{f} (\mathbf{r}, \mathbf{x}, \Delta) =       
    \begin{cases}
          1/\Delta^3  & \text{for } |\mathbf{r}| \le \Delta /2, \\    
          0  & \text{otherwise\, .}  
            \end{cases}
\label{eq:tophat}
\end{equation}

The continuity and momentum equations for the incompressible fluid flow, with LES filtering applied, can be written as

\begin{equation}
\frac{\partial \overline{u}_i} {\partial x_{i}}=0, \, ~~ \frac{\partial \overline{u}_i}{\partial t}+\frac{\partial}{\partial x_j}(\overline{u_i u_j})=-\frac{1}{\rho} \frac{\partial \overline{p}}{\partial x_i}+\nu \frac{\partial^2 \overline{u}_i}{\partial x_j \partial x_j},
\label{nsfiltered1}
\end{equation}

\noindent where $\overline{u}_i$ is the filtered velocity, $\overline{p}$ represents the filtered static pressure and $\nu$ is the kinematic molecular viscosity. The term $\overline{u_i u_j}$ is the dyadic product and cannot be expressed directly \citep{Ferziger-NMFM1998}. Modification of the momentum equation (\ref{nsfiltered1}) by $+\frac{\partial}{\partial x_j}(\overline{u}_i \overline{u}_j)$ yields

\begin{equation}
\frac{\partial \overline{u}_i}{\partial t}
+\frac{\partial}{\partial x_j}(\overline{u}_i\overline{u}_j)=
-\frac{1}{\rho} 
\frac{\partial \overline{p}}{\partial x_i}+\nu \frac{\partial^2 \overline{u}_i}{\partial x_j \partial x_j}-\frac{\partial \tau_{ij}}{\partial x_j} .
\label{eqn:filtered-ns-2}
\end{equation}

\noindent The new term on the right-hand side of \eqref{eqn:filtered-ns-2} is the divergence of the subgrid-scale turbulent stress tensor

\begin{equation}
\tau_{ij}=\overline{u_i u_j} - \overline{u}_i \overline{u}_j=- ( \overline{u_{i}' u_{j}'} + \overline{\overline{u}_i u_{j}'} + \overline{u_{i}' \overline{u}_j }  +  \overline{\overline{u}_i  \overline{u}_j }  -  \overline{u}_i \overline{u}_j )
\label{eq:tauije}
\end{equation}

\noindent and left to be modeled to close the set of equations. Since the turbulence is not fully understood, a wide range of closure subgrid-scale models have been introduced, often using heuristic and ad-hoc techniques.

\subsubsection{One-equation SGS model} 

The OE model derived by \citet{Yoshizawa-PSJ1985} computes the transport equation for the turbulent kinetic subgrid-scale energy $k_{SGS}$

\begin{equation*}
\frac{\partial k_{SGS}}{\partial t}+\frac{\partial \overline{u}_j k_{SGS}}{\partial x_j} - \frac{\partial}{\partial x_j} \left[(\nu+\nu_t) \frac{\partial k_{SGS}}{\partial x_j} \right] = 
\end{equation*}
\begin{equation}
-\tau_{ij} \overline{S}_{ij} - C_{\epsilon} \frac{k_{SGS}^{3/2}}{\Delta},
    \label{eq:oe0}
\end{equation}

\noindent \deleted{The model constant in \eqref{eq:oe0} is set to $C_{\epsilon}=1.048$} \added{{where $C_{\epsilon}=1.048$ is model constant, $\overline{S}_{ij}$ is resolved rate-of-strain tensor (symmetric part of the velocity-gradient tensor) and $\nu_{t}$ is turbulent viscosity.}}

\noindent Unlike the Smagorinsky model \citep{Smagorinsky-MWR1963}, which disregards the first three terms in \eqref{eq:oe0}, the OE model considers also the history effects for $k_{SGS}$. The production term $-\tau_{ij} \overline{S}_{ij}$, modeling the decay of turbulence from the resolved scales to the SGS scales via the energy cascade, is approximated by

\begin{equation} 
 -\tau_{ij} \overline{S}_{ij} = 2 \nu_{t} \overline{S}_{ij}\overline{S}_{ij}.
\end{equation}

\replaced{The one-equation}{OE} model relies on the SGS eddy viscosity

\begin{equation}
 \quad \nu_t^{O}=C_{\nu} \Delta \sqrt{k_{SGS}},
\label{eqn:nut-oneeq}
\end{equation}

\noindent where $C_{\nu}=0.094$, due to the Kolmogorov law.

Neither the Smagorinsky nor \replaced{the One-equation}{OE model} can reproduce the laminar to turbulent transition and tend to overpredict the production rate and thus the turbulent viscosity in free shear layers and near the walls. This is caused by the fact that the term $\overline{S}_{ij}\overline{S}_{ij}$ is large in the regions of pure shear because it is only related to the rate-of-strain $\overline{S}_{ij}$, not to the rate-of-rotation $\overline{\Omega}_{ij}$ \citep{Lesieur-book2005}.
 
  \subsubsection{WALE SGS model}
   
   The inaccuracy concerning free shear and boundary layer treatment, caused by the OE model, can be alleviated by using the WALE model \citep{Nicoud-FTC1999}. \deleted{The WALE model considers the traceless symmetric part of the square of the velocity gradient $s_{ij}^{d}$ written as}


\noindent \deleted{ where $\overline{G}_{ij}^2=\overline{G}_{ik}\overline{G}_{kj}$, $\overline{G}_{ij}=\partial \overline{u}_i / \partial x_j$ is the filtered velocity gradient tensor, $\delta_{ij}$ is the Kronecker symbol. }


\noindent \deleted{where the terms are rate-of-strain $\overline{S}_{ij}$ and rate-of-rotation $\overline{\Omega}_{ij}$ tensors}





\noindent \deleted{is used. The term (.)} \added{The WALE model} is able to detect turbulent structures with strain, or rotation rate, or even both. Regarding this behavior, the pure shear flow located near solid boundaries during laminar flow will cause that the eddy-viscosity vanishes \citep{Nicoud-FTC1999}.

The turbulent viscosity computed by the WALE model is defined as

\begin{equation}
    \nu_{t}^{W}= C_k \Delta \sqrt{k_{SGS}^{W}},
    \label{eq:nutwof}
\end{equation}

\noindent where the term $k_{SGS}^{W}$ is

\begin{equation}
    k_{SGS}^{W}= \left( \frac{C_{w}^2 \Delta}{C_k} \right)^{2} \frac{(s_{ij}^ {d} s_{ij}^ {d} )^{3}}{ \left( (\overline{S}_{ij}\, \overline{S}_{ij})^{5/2} + (s_{ij}^ {d} s_{ij}^{d} )^{5/4} \right)^2 }.
    \label{eq:ksgswof}
\end{equation}

\noindent The model constants are set to $C_w=0.325$ and $C_k=0.094$, \added{and $s_{ij}^{d}$ is the traceless symmetric part of the square of the velocity gradient}.

In summary, the~WALE algebraic model formulation accounts for the rotational rate in the computation of $\nu_{t}^{W}$, and thus the turbulent viscosity tends to zero near walls. Hence, it is not necessary to use any \textit{ad hoc} damping methods.

  \subsubsection{AMD SGS model}

The anisotropic minimum-dissipation (AMD) subgrid-scale model was derived by Rozema \citep{Rozema-PF2015} with modified Poincaré inequality addressing the grid anisotropy. \deleted{AMD is developed from the QR model $\citep{Verstappen-JSC2011}$, and both models are in}\added{The model belongs to the category "minimum-dissipation models"}. The main objective of these models is to ensure that the energy of subgrid scales $k_{SGS}$ is not increasing

\begin{equation}
    \partial_t \int_{\Omega_{\Delta}}^{} \frac{1}{2} u_{i}' u_{i}' {\rm d}x \leq 0.
\end{equation}

\noindent In the situation where subgrid scales are assumed to be periodical on the filter box $\Omega_{\Delta}$, it is possible to apply the Poincaré inequality, and define thus the upper bound of $k_{SGS}$

\begin{equation}
    \int_{\Omega_{\Delta}} \frac{1}{2} u_{i}' u_{i}'  {\rm d}x \leq C \int_{\Omega_{\Delta}} \overbrace{\frac{1}{2} ( \partial_i u_j ) ( \partial_i u_j )}^{R1} {\rm d}x.
    \label{eq:poincareQR}
\end{equation}

\noindent The term $R1$ in \eqref{eq:poincareQR} corresponds to the velocity gradient energy, and $C$ is the Poincaré constant $C=(\Delta / \pi)^2$ for the LES filter of width $\Delta$.

The AMD model can sidestep the dependence of the model constant on $\Delta$ by using the modified Poincaré inequality

\begin{equation}
    \int_{\Omega_{\Delta}} \frac{1}{2} u_{i}' u_{i}'  {\rm d}x \leq C_{A} \int_{\Omega_{\Delta}} \overbrace{\frac{1}{2} (\underbrace{\Delta x_i \partial_i}_{R3} u_j ) (\Delta x_i \partial_i u_j )}^{R2} {\rm d}x,
    \label{eq:poincare}
\end{equation}

\noindent where \replaced{$\Omega_{\delta}$}{$\Omega_{\Delta}$} is the filter box, having dimensions $\Delta x_1$, $\Delta x_2$ and $\Delta x_3$, and $C_{A}$ is the modified Poincaré model constant\deleted{, which will be discussed later}. The term $R2$ is the scaled velocity gradient energy, $R3$ the scaled gradient operator. The inequality \eqref{eq:poincare} demonstrates that the subgrid energy is confined by imposing a bound on the term $R2$ \citep{Rozema-PF2015}. Time derivative is applied on the term $R2$ and the evolution equation of $R2$ on the filter box $\delta x_i$ is expressed

\begin{equation*}
\partial_t \bigg( \frac{1}{2} (\Delta x_i \partial_i u_j) (\Delta x_i \partial_i u_j \bigg)=\overbrace{-(\Delta x_k \partial_k u_i) (\Delta x_k \partial_k u_j) S_{ij}}^{R4}  
    \end{equation*}

\begin{equation}
    - (\nu+\nu_{t}^{A}) \Delta x_k \partial_k (\partial_i u_j) \Delta x_k \partial_k (\partial_i u_j)+\partial_i f_i,
    \label{eq:amdd}
\end{equation}

\noindent where the term $R4$ is the production of the scaled velocity gradient energy. The following inequality in \eqref{eq:poincare3} ensures that the AMD model predicts sufficient dissipation to stop the production of scaled velocity gradient energy $R4$

\begin{equation}
    \int_{\Omega_{\Delta}} R4~  {\rm d}x \leq \frac{\nu_{t}^{A}}{C_A} \int_{\Omega_{\Delta}} (\partial_i u_j) (\partial_i u_j) {\rm d}x,
    \label{eq:poincare3}
\end{equation}

\noindent where the minimum dissipation effect is held by satisfying

\begin{equation}
\nu_{t}^A = C_A \frac{{\rm max} \{ \int_{\Omega_{\Delta}}-(\Delta x_k \partial_k u_i) (\Delta x_k \partial_k u_j) S_{ij}  {\rm d} x, 0 \}}{\int_{\Omega_{\Delta}} (\partial_l u_m) (\partial_l u_m) {\rm d}x}.
    \label{eq:poincare4}
\end{equation}

\noindent Integrals in \eqref{eq:poincare4} can be approximated by the mid-point rule, and the turbulent viscosity from AMD $\nu_{t}^A$ results in a more practical form

\begin{equation}
\nu_{t}^A = C_A \frac{ {\rm max} \{ - \overbrace { (\delta x_k \partial_k u_i) (\delta x_k \partial_k u_j) S_{ij}}^\text{$R4$}, 0 \} }{ \underbrace{(\partial_l u_m) (\partial_l u_m)}_{R5}},
    \label{eq:poincare5}
\end{equation}

\noindent where the terms in vector notations are

\begin{equation}
    R4  =   (\Delta \mathbf{x} \nabla \mathbf{u}) \cdot (\Delta \mathbf{x} \nabla \mathbf{u}^\top): \mathbb{{S}}, \,~~ R5 = (\nabla \mathbf{u}) : (\nabla \mathbf{u}).
\end{equation}

\added{Implementation of AMD is available  into OpenFOAM in the AMD library\footnote{https://gitlab.com/mlasota/myFoam}}

The constant $C_A$ suitable for the AMD model is recommended from \citep{Rozema-PF2015} with respect to the order of discretization of Navier-Stokes equations, tested on decaying grid turbulence cases\replaced{.}{, and Rozema has concluded that the} AMD model gave the best results with $C_{A}=0.3$ for a central second-order scheme and $C_A=0.212$ for a fourth-order scheme. A recent study by \citep{Zahiri-CF2019} states an optimal value of the constant $C_A=\frac{1}{\sqrt{3}}=0.577$ based on various study cases. \citet{Lasota-DISS2022} performed own tests of model constants ($C_A=0.3$ and $C_A=0.57735$) on cases with turbulent plane channel and periodic hill\deleted{, based on this $C_A=0.3$ was chosen in this study}\added{. Based on better agreement with velocity profiles obtained by direct numerical simulation or experiments, $C_A=0.3$  was chosen as the model constant for turbulence modeling in the larynx}.

\citet{Rozema-PF2015} has shown that after a Taylor expansion of $\tau_{ij}$ in \eqref{eq:tauije},\deleted{that the AMD model is really consistent with the exact subgrid-scale stress tensor} \added{consistency between the AMD model and the exact subgrid-scale stress tensor $\tau_{ij}$ can be proved. After Taylor expansion of the eddy dissipation of the exact subgrid-scale stress tensor $\tau_{ij} S_{ij}$, consistency with $R5$ in \eqref{eq:poincare5} can be shown.}

\deleted{
\noindent and the eddy dissipation of the exact subgrid-scale stress tensor is approximated as }

\noindent \deleted{which means that the term $R5$ in (.) is consistent with the product of Taylor series in (.).} \deleted{The term $R5$ is also referred as the gradient sub-filter model.} If the exact eddy dissipation gives zero dissipation, then the term $R5$ as well, this means the AMD model can be switched off for flows where the exact eddy dissipation is vanishing. Thus, the AMD model also switches off when no subgrid energy is created \citep{Rozema-PF2015,Vreugdenhil-PF2018}.

 \subsection{Geometry and boundary conditions}

\begin{figure}[ht]
    \centering
    \includegraphics[width=1\linewidth]{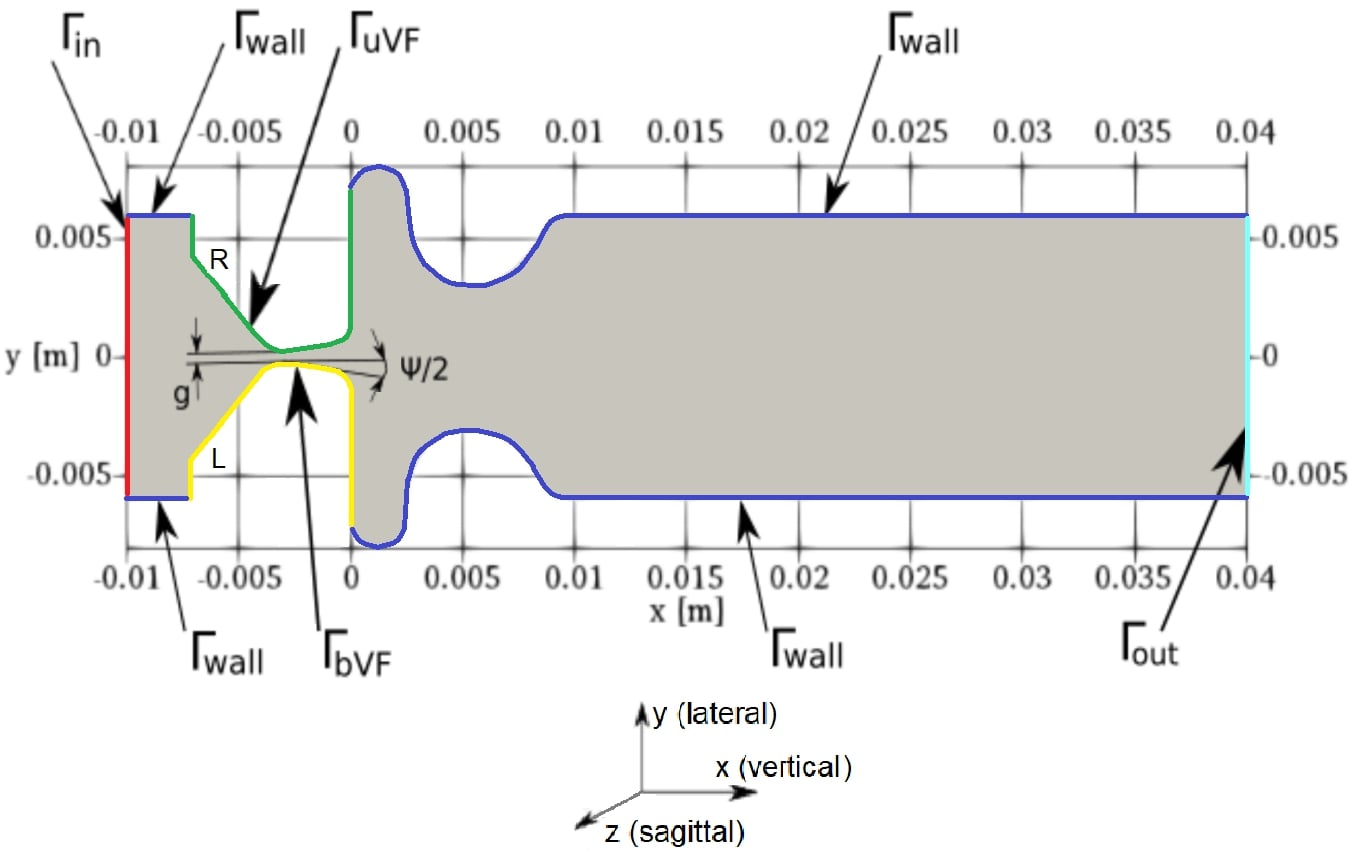}
    
\caption{Mid-coronal (x-y) section of the CFD computational domain. The z-normal (front and back) boundaries belong to $\Gamma_{{\rm wall}}$.}
    \label{fig:geo}
\end{figure}

\noindent The geometry of vocal folds is based on the M5 parametric shape by \citep{Scherer-JASA2001}, (see Fig.~\ref{fig:geo}). The false vocal folds were specified according to data published by \citep{Agarwal-JV2003}. The geometrical model is in 3D, having a square cross-section at inlet 12x12 mm. More details can be found in \citep{Sidlof-BMM2015}. 

The boundary conditions for the CFD model are summarized in Tab.~\ref{fig:bcs}. The flow is driven by constant pressure difference $P_k=\overline{p}/\rho=300~{\rm m^2/s^s}$ between the inlet $\Gamma_{{\rm in}}$ and outlet $\Gamma_{{\rm out}}$. The velocity on $\Gamma_{{\rm in}}$ and $\Gamma_{{\rm out}}$ is computed from the flux. \added{Turbulence initialization at inlet is not used since turbulent intensity upstream of the glottis is low and the flow at inlet can be considered laminar \citep{Lasota-TPFM2019}.}

\begin{table}[ht]

\caption{Boundary conditions for the filtered flow velocity $\mathbf{\overline{u}}$ and static pressure $\overline{p}$. The symbol $\mathbf{n}$ is the unit outer normal and $ h(\mathbf{x},t)$ is the prescribed displacement of the vocal folds.}
\vspace{0.1cm}
\centering
\begin{tabular}{l|l|l}
\textbf{}	& $\mathbf{\overline{u}} \ \rm [ms^{-1}]$	&  $\overline{p} \ \mathrm{[Pa]}$ \\
\hline
  $\Gamma_{{\rm in}}$  & $\mathbf{\overline{u}}=\mathbf{0}$~if $\mathbf{\overline{u}} \cdot \mathbf{n} < 0,$ & 350\\ 
    & $\nabla \mathbf{\overline{u}} \cdot \mathbf{n}=\mathbf{0}$~if~$\mathbf{\overline{u}} \cdot \mathbf{n} > 0 $ & {~} \\ 
\hline
   $\Gamma_{{\rm out}}$ & $\mathbf{\overline{u}}=\mathbf{0}$~if~$\overline{\mathbf{u}} \cdot \mathbf{n} <0,$ & 0 \\
   {~} & $\nabla \mathbf{\overline{u}} \cdot \mathbf{n}=\mathbf{0}$~if~$\overline{\mathbf{u}} \cdot \mathbf{n} > 0$ & {~} \\
 \hline
  $\Gamma_{{\rm bVF}}$  & $\overline{u}_1 = 0$, $\overline{u}_2=\frac{\partial}{\partial t} h(\mathbf{x},t)$ &  $\nabla \overline{p} \cdot \mathbf{n}=0$  \\

    & $\overline{u}_3 = 0$ &    \\ 
  
   \hline
  $\Gamma_{{\rm uVF}}$   & $\overline{u}_1 = 0$, $\overline{u}_2=\frac{\partial}{\partial t} h(\mathbf{x},t)$ &  $\nabla \overline{p} \cdot \mathbf{n}=0$  \\ 
  
      & $\overline{u}_1 = 0$, $\overline{u}_3 = 0$ &    \\ 
  
   \hline
     $\Gamma_{{\rm wall}}$ & $\mathbf{\overline{u}=0}$ &  $\nabla \overline{p} \cdot \mathbf{n}=0$   \\
     
\end{tabular}
\label{fig:bcs}
\end{table}

\noindent On the moving boundaries $\Gamma_{{\rm bVF}}$ and $\Gamma_{{\rm uVF}}$, the flow velocity is equal to the velocity of the moving vocal fold surface, given by the function $h(\textbf{x},t)$. The function $h(\mathbf{x},t)$ based on the sinusoidal displacement~$w_{1,2}= A_{1,2} \sin (2\pi f_o t + \xi_{1,2})$ ensures the vibrating motion of vocal folds in the medial-lateral ($y$) direction with two degrees of freedom. In the current simulation, the vocal folds oscillate symmetrically with a frequency $f_o$ = 100~Hz, amplitudes at the superior and inferior vocal fold margin are $A_{1} = A_{2} = 0.3~\rm mm$. The medial surface convergence angle is marked in Fig.~\ref{fig:geo} as $\psi/2$, which confines the convergent and divergent position (-10~$\deg$ and +10~$\deg$). The distance (y) between both ventricles and both false vocal folds equals 16 and 6.15~mm, respectively. In this study, the oscillation of the vocal folds allows closing/opening the glottal gap $g$ in the range 0.42-1.46 mm.

 \subsection{Mesh and discretization }
 In wall-bounded flows, the presence of solid walls fundamentally influences the flow dynamics, turbulence generation, and transport in the near-wall regions due to significant viscous stresses. The accuracy of the numerical simulation is thus closely related to the grid resolution near the fixed walls. According to the classification by \citet{Pope-book2000}, large-eddy simulations of wall-bounded flows can be classified as large-eddy simulations with near-wall resolution (LES-NWR) with a grid sufficiently fine to resolve 80\% of the turbulent energy in the boundary layer, and large-eddy simulation with near-wall modeling (LES-NWM), which employs a modeling approach similar to \added{Reynolds-Averaged Navier-Stokes} (RANS) in the near-wall region. For these simulations, an important parameter is the wall unit $ y^+ = u_\tau y/\nu $, where $u_{\tau}=\sqrt{|\tau_{w}|/\rho}$ is the friction velocity, $\tau_w = \mu_{\mathrm{eff}} \left( \partial U/\partial y \right) \Big|_{y=0}$ is the wall shear stress, $\mu_\mathrm{eff}=\left( \mu + \mu_t \right)$ is the effective dynamic viscosity and $y$ the dimensional distance in normal direction from the wall.

Using the same normalization, $x^+$ and $z^+$ denote the dimensionless streamwise and spanwise distances. Wall units are also commonly used to indicate LES adequacy. According to \citep{Georgiadis-AIAAJ2010} and \citep{Jiang-book2016}, in LES-NWR the theoretical limits for the grid spacing adjacent to the wall are $50 \le \Delta x^{+} \leq 150$, $\Delta y^{+}< 1$ and $15 \le \Delta z^{+} \leq 40$, with at least 3-5 gridpoints between $0 < y^+ < 10$. 

The computational mesh in the current CFD simulation is block-structured to capture well the boundary layer and consists of 2.2M hexahedral elements. An open-source 3D finite volume mesh generator, blockMesh \added{(part of the OpenFOAM)} was used to build the mesh. The mesh deforms in time due to vocal fold oscillation. The grid resolution in wall units was evaluated in three distinct time instants, corresponding to a maximum opening of the vocal folds, full closure during the divergent phase and full closure during the convergent phase. On the boundary $\Gamma_{{\mathrm{bVF}}}$ at the critical time when the vocal folds are maximally adducted were evaluated these values: $y^{+}_{avg}=1.77$, $z^{+}_{}=14$ and $x^{+}_{}=8$.

The Navier-Stokes equations were discretized using the collocated cell-centered Finite Volume Method. Fletcher \citep{Fletcher-FD1991} demonstrated that even-ordered derivatives in the truncation error are associated with numerical dissipation, and odd-ordered spatial derivatives are associated with the numerical dispersion in the solution. Ideally, LES simulations should use schemes with low numerical dissipation. The non-dissipative central differencing scheme, which was applied in this study, allows an accurate representation of the changing flow field \citep{Launchbury-book2016}. The discretization of the diffusion term is split into an orthogonal and cross-diffusion term, using a procedure described in \citep{Jasak-DISS1996}. Unlike the discretization of the temporal, convective, and orthogonal part of the diffusive term, the non-orthogonal correctors are treated explicitly.
 
 \subsection{Numerical solution}

CFD simulations were run in parallel on either \textit{Charon} (Metacentrum NGI - Technical University of Liberec, 20 cores on a computational cluster, composed of nodes with two 10-core Intel Xeon Silver 4114 2.20GHz CPUs with 96GB RAM) or 
\textit{Fox} (Computing center of the Czech Technical University in Prague, 20 cores on a supercomputer (SGI Altix UV 100) with shared memory 576GB RAM with the involvement of 6-core Intel Xeon Nehalem 2.66GHz CPUs).

In order to have sufficient resolution in the spectrum of the aeroacoustic signal, a sufficiently long simulation time $t = 0.2~{\rm s}$, i.e., 20 periods of vocal fold vibration, is needed. For such a setting, one CFD simulation required 27 - 37 days, i.e., about 15000 core-hours of computational time.

 \subsection{CFD results}
 
This work reports on the results of CFD simulations using different turbulence modeling approaches, which are summarized in Tab.~\ref{cases2}. \added{The CFD model was validated on two benchmarks (backward-facing step, periodic hill) \citep{Lasota-DISS2022}.}

\begin{table}[ht]
    \caption{Overview of the CFD simulations.}
    \vspace{0.1cm}
    \centering
    \begin{tabular}{cccccc}
        \textbf{Case} & \textbf{Type} & \textbf{SGS} &  \textbf{Cluster}  & \textbf{Wall-time} \\ 
        \hline 
        LAM	& laminar & - &  Charon  & 27d~13h\\
        OE & LES & OE &  Charon & 34d~05h\\
        WALE & LES & WALE &  Charon & 37d~13h \\
        AMD & LES & AMD & Fox & 34d~18h \\
        \hline 
    \end{tabular}
    \label{cases2}
\end{table}

  \subsubsection{Laryngeal flow rate}
\begin{figure}[ht]
    \centering
      \includegraphics[width=1\linewidth]{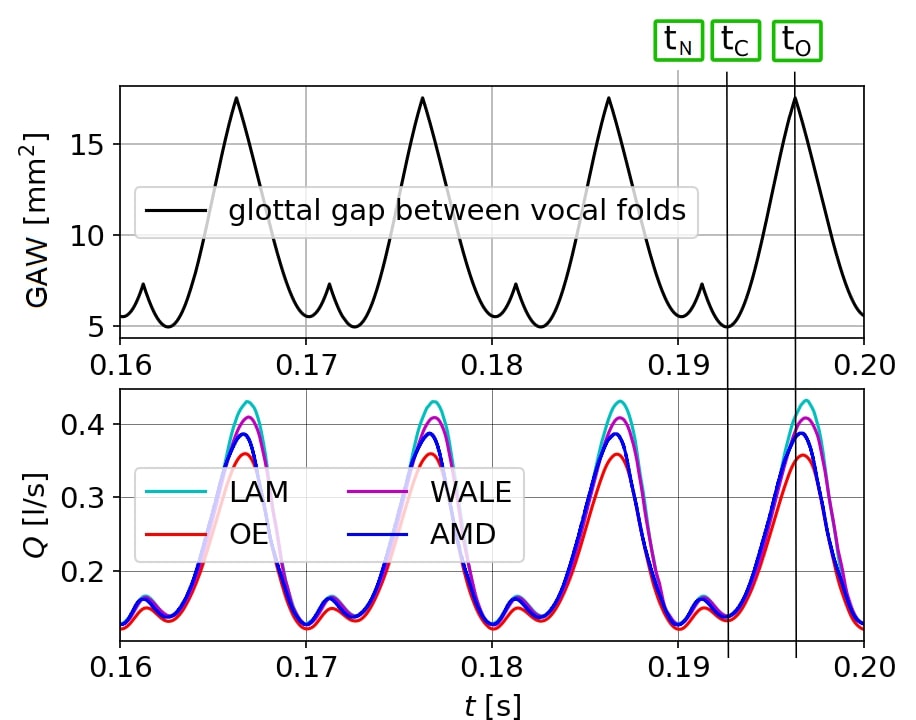}
    \caption{Glottal area waveforms (GAW) and flow rates during four oscillation cycles. Time instants for further analysis: $t_{\mathrm{N}}=0.1900~\mathrm{s}$, $t_{\mathrm{C}}=0.1927~\mathrm{s}$ and $t_{\mathrm{O}}=0.1963~\mathrm{s}$.}
    \label{QPUx}
\end{figure}

  Fig.~\ref{QPUx} shows the glottal opening and flow rates during the last four simulated cycles of vocal fold oscillation. The time $t_N$ corresponds to the instant where the inferior margins of the vocal folds approach most and reduce the glottal opening to $5.58~{\rm mm^2}$ ($g=0.465~{\rm mm}$). Time instant $t_C$ is the maximum approach of the superior vocal fold margins, where the glottal opening drops to $4.98~{\rm mm}^2$ ($g=0.415~{\rm mm}$). The third time instant, $t_O$, corresponds to the maximum glottal opening of $17.51~{\rm mm}^2$ ($g=1.459~{\rm mm}$).
  
\noindent The subgrid-scale models affected the flow rates $Q$[l/s] (see Fig.~\ref{QPUx}): the predicted peak flow rate in the laminar case is higher than in the One-equation, WALE and AMD SGS models by 16.76\%, 5.26\% and 9.3\%, respectively. This is caused by the different values of the SGS viscosity, which adds to the molecular viscosity and limits the flow rate through the glottal constriction. The laminar model does not capture the influence of small-scale turbulence, which corresponds to $\nu_{t} = 0$. The WALE SGS model and the One-equation SGS model compute with non-zero SGS viscosity, with the latter one significantly higher due to the already mentioned deficiency of the One-equation model, which overestimates the turbulent viscosity near the vocal fold surfaces. In all cases, the flow rate does not reach zero value, corresponding physiologically to breathy phonation. The vocal folds do not fully close the glottal channel from technical reasons. The minimum flow rate is $Q_{min}\approx 0.122~{\rm l/s}$ by using OE. The maximum flow rates are between $0.358-0.434~{\rm l/s}$ by using WALE and LAM models. The peak flow rate predicted by the AMD model occurs sooner than in other simulations (when 66\% of the VF cycle is reached).

  \subsubsection{Vorticity field}
Vorticity ($\boldsymbol{\omega}=\nabla \times \mathbf{u}$) is commonly used for characterizing turbulent structures in cases with no entrainment rotation. The vorticity fields reveal the shear layers, where vortices are shed as a consequence of Kelvin-Helmholtz instability. The vortices may undergo successive instabilities, leading to a direct kinetic-energy cascade towards the small scales.

\begin{figure}[ht]
    \centering
    \includegraphics[width=1\linewidth]{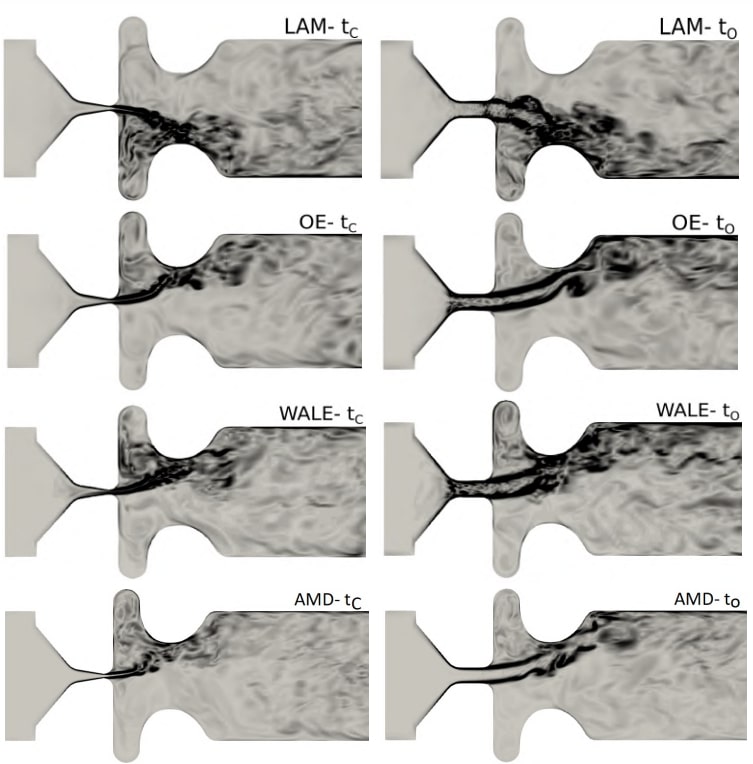}
    \caption{Vorticity fields $\left| \boldsymbol{\omega} \right| $ in mid-coronal plane in range (0,30000) [${\rm s^{-1}}$].}
    \label{fig:vorticity_all}
\end{figure}

Fig.~\ref{fig:vorticity_all} shows vorticity fields presented in mid-coronal plane (x-y). The supraglottal jet deflects stochastically towards either of the ventricular folds. This behavior is not a consequence of the SGS model, it is caused by the bistability of the flow in this symmetric geometry \citep{Erath-EF2010,Lodermeyer-EF2015}. Detailed analysis of the vorticity within the glottal region shows that the average value of vorticity in glottal region is similar for all SGS models. 


\begin{figure}[ht]
    \centering
        \includegraphics[width=1\linewidth]{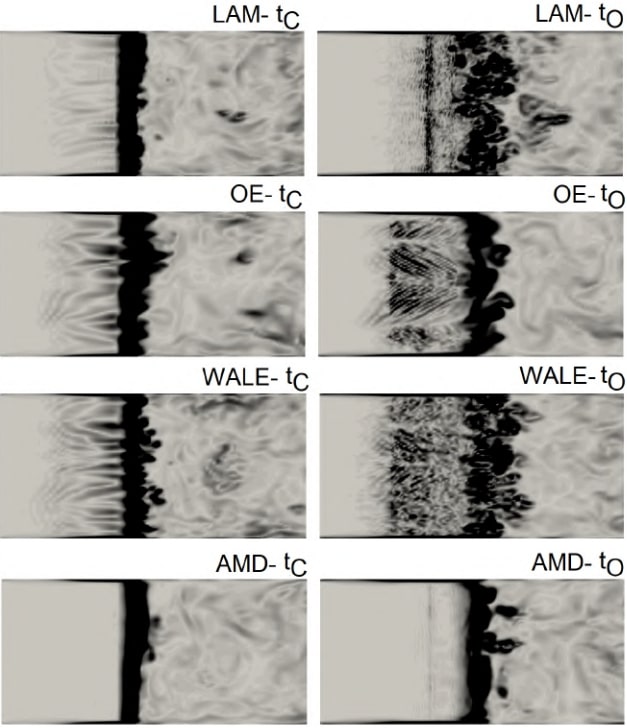}
        \caption{Vorticity fields $\left| \boldsymbol{\omega} \right| $ in mid-sagittal plane in range (0,30000) [${\rm s^{-1}}$].}
    \label{fig:vorticity_xz}
\end{figure}

Fig.~\ref{fig:vorticity_xz} shows a complementary view on the magnitude of the vorticity vector $\left| \boldsymbol{\omega} \right| $ in mid-sagittal plane (x-z). The simulation with the AMD model predicts low vorticity near the glottis. The absence of vorticity may imitate the situation in the realistic larynx where the jet is frequently stopped and renewed, and thus the turbulent eddies are forced to be dissipated.

  \subsubsection{Turbulent viscosity field}
  
The effect of the unresolved turbulent subgrid scales on the resolved scales is carried by the subgrid-scale turbulent viscosity $\nu_t$, represented by equations \eqref{eqn:nut-oneeq} for $\nu_{t}^{O}$, \eqref{eq:nutwof} for $\nu_{t}^{W}$ and \eqref{eq:poincare5} for $\nu_{t}^{A}$.

\begin{figure}[ht]
  \centering
 \includegraphics[width=1\linewidth]{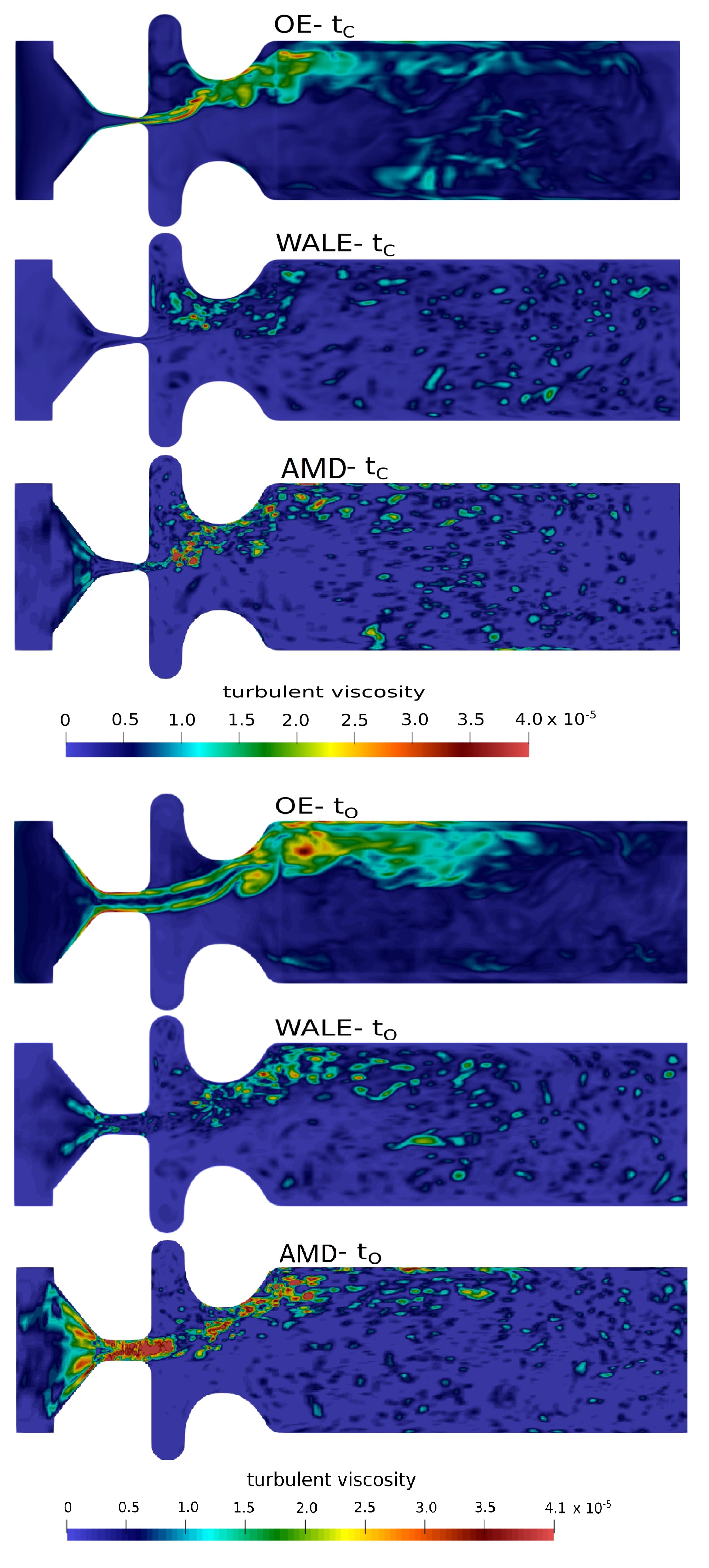} 
\caption{Turbulent viscosity $\nu_t$ ${\rm[m^2.s^{-1}}]$ in mid-coronal plane at $t_C$ and $t_O$.}
    \label{fig:viscosity_all0}
\end{figure}

Fig.~\ref{fig:viscosity_all0} shows that the turbulent viscosity predicted by the simulation with the One-equation model is very high in regions of pure shear, especially within glottis. This may be the reason the simulation with the OE model predicted usually the lowest intraglottal velocity. In contrast to this, WALE and AMD subgrid-scale models predicted considerably lower subgrid-scale viscosity in the shear layers at $t_{\mathrm{C}}$. The fields computed by the AMD model seem to be similar to fields computed by WALE with spots of gently higher subgrid-scale viscosity at $t_{\mathrm{C}}$. The other situation occurs in $t_{\mathrm{O}}$ when the turbulent viscosity predicted by the AMD model is around two times higher than by OE and five times higher than by WALE.

\begin{figure}[ht]
    \centering
        {\includegraphics[width=1\linewidth]{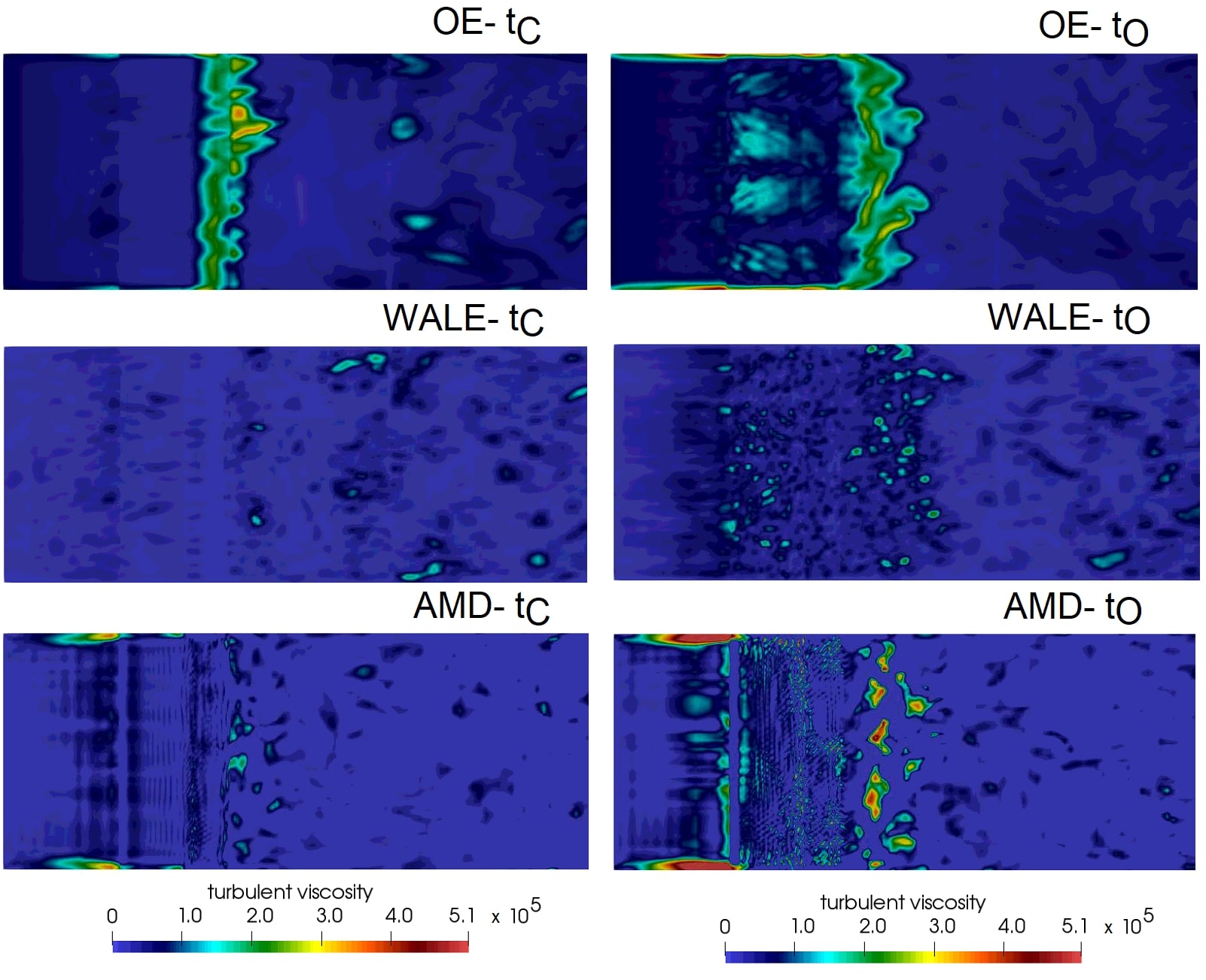}}
            
    \caption{Turbulent viscosity $\nu_t$ $\rm[m^2.s^{-1}]$ in mid-sagittal plane at $t_C$ and $t_O$.}
    \label{fig:nutxz1}
\end{figure}

Fig.~\ref{fig:nutxz1} shows turbulent viscosity fields in the mid-sagittal plane. The simulation with OE predicted twice higher turbulent viscosity located at the vicinity of the inferior margin of the vocal folds than others. The narrow barrier of turbulent viscosity at $t_N$ in the case with AMD reduced the flow rate just by 0.8\% (1 ml/s of air) compared to WALE.

\section{CAA model} \label{Chap:CAA}

 \subsection{Mathematical model}

  \subsubsection{Perturbed convective wave equation (PCWE)}
  
In this article, the PCWE \added{aeroacoustic model} \citep{Hueppe-AIAA2014} was used in all cases. If the overbars ($\overline{\cdot}$) from the LES filtering are dropped for simplicity, \added{the governing equation for the acoustic potential $\psi^{a}$ is}

\begin{equation}
    \frac{1}{ c_{0}^2} \frac{{\rm D}^2 \psi^{a}}{{\rm D}t^2} - \nabla \cdot \nabla(\psi^{a}) = - \frac{1}{\rho_0 c_{0}^2} \frac{{\rm D}p^{ic}}{{\rm D}t},
    \label{050520-2018}
\end{equation}

\noindent \added{where $c_0$ is speed of sound, $\mathrm{D}/\mathrm{D}t$ is total derivative operator, $\rho_0$ is ambient density, $p^{ic}$ is incompressible pressure.} The following should be remembered: 

\begin{itemize}
    \item The sound generation \citep{Schoder-APPSCI2021} is discussed in terms of the \replaced{material}{total temporal} derivative \citep{Schoder-AIAA2022} of the incompressible fluid dynamic pressure obtained from the CFD part, i.e. ${\rm D}p^{ic}/{\rm D}t$, in dimension [Pa/s].
        \item The wave propagation \eqref{050520-2018} contains the factor $-1/(\rho_0  c_{0}^2)=-1/(1.1493 \cdot 351.31^2)=-0.000007~{\rm ms^2/kg}$. The parameters \added{$(\rho_0, c_{0})$} were chosen taking into account the physiological human breathing temperature of $34\,^\circ{\rm C}$ \citep{Anghel-ICM2013}.  
        \item From the solution of equation \eqref{050520-2018}, i.e., acoustic potential $\psi^a$, the acoustic pressure $p^{{a}}$ [Pa] is calculated according to
        
        \begin{equation}
    p^{a} = \rho_{0} \frac{{\rm D} \psi^{a}}{{\rm D}t} =\rho_{0} \frac{\partial \psi^a}{\partial t} + \overbrace{\rho_{0} \mathbf{u}_0 \cdot \nabla \psi^{a}}^{\approx0},
    \label{040520-1406}
\end{equation}
        
        \deleted{where the convective term $\rho_{0} \mathbf{u}_0 \cdot \nabla \psi^{a}$ contributes only a minor part to the solution $\citep{Schoder-APPSCI2021}$, and thus $p^a=\rho_{0} \partial \psi^a / \partial t$ is computed.}\added{where the convective term $\rho_{0} \mathbf{u}_0 \cdot \nabla \psi^{a}$ contributes only a minor part to the solution (three orders of magnitude smaller than the other terms), and thus only $p^a=\rho_{0} \partial \psi^a / \partial t$ is used} \citep{Schoder-APPSCI2021}.
\end{itemize}

\noindent Benefits of PCWE are as follows: The vector-valued \added{acoustic perturbation equation (APE-2) for isothermal and low Mach numbers \citep{Hueppe-AIAA2014}} is reformulated into the scalar PCWE, which can save computational resources avoiding the vector form. That means faster computation with a scalar unknown, lower memory requirements compared to APE-2, includes convection inside the wave operator and solves the acoustic quantity compared to Lighthill's analogy \citep{Lighthill-PRSL1952}. Finally, the PCWE can be extended to account for moving vocal folds \citep{Schoder-ARXIV2022}.

 \subsection{Geometry, mesh and boundary conditions}
 
\begin{figure}[ht]
    \centering
  \includegraphics[width=1\linewidth]{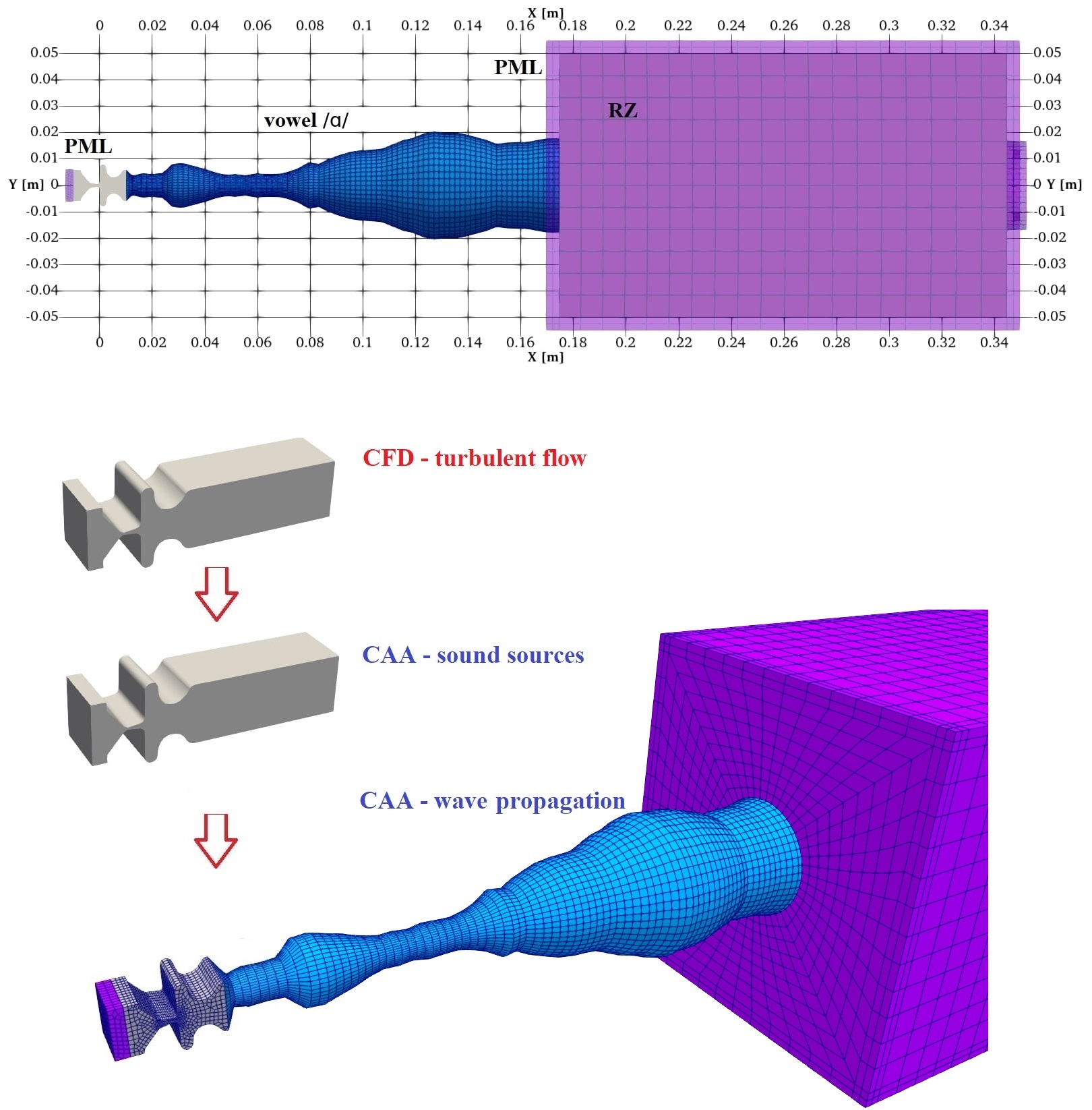}
\caption{Computational domain for the aeroacoustic simulation composed of the perfectly matched layer (PML), vocal folds, false vocal folds, vocal tract and radiation zone (RZ) covered by PML. Arrows show the workflow; sound sources were computed from the CFD results and visualised on the CFD mesh in high-resolution, and afterwards sound sources are interpolated to the coarse "CAA-wave propagation" mesh to perform the acoustic simulation.}
\label{fig:colors}
\end{figure}

 Fig.~\ref{fig:colors} illustrates one of the geometry used for aeroacoustic simulations, where from the left side are: the PML \added{(Perfectly Matched Layer)} \added{with a thickness of 0.3 cm at inlet}, larynx with vocal folds \added{and false vocal folds (2 cm)}, vocal tract \added{(vowels /\textipa{A}, o/: 17.46 cm, /u/: 18.25 cm, /i, æ/: 16.67 cm)} and the radiation zone \added{(RZ)} protected by the \added{2-3} PMLs \deleted{hollow cube} \added{with a thickness of 0.5 cm}. PML \deleted{(Perfectly matched layer)} is a technique published first by \citep{Berenger-JCP1994}; the original method was modified by \citep{Kaltenbacher-JCP2013} by adding damping layers to guarantee that no wave reflections occur at boundaries.

\noindent The geometry of the vocal tract was modeled from frustums (0.397~{\rm cm} long) concatenated one after another. The shape of the frustums was defined according to the vocal tract area function measured by magnetic resonance imaging \citep{Story-JASA1996}. \added{The first three frustums were removed from the model otherwise the false vocal folds would be there twice}. \added{About 33k hexahedral first-order finite elements were used for each vowel.}

\begin{figure}[ht]
    \centering
      \includegraphics[width=1\linewidth]{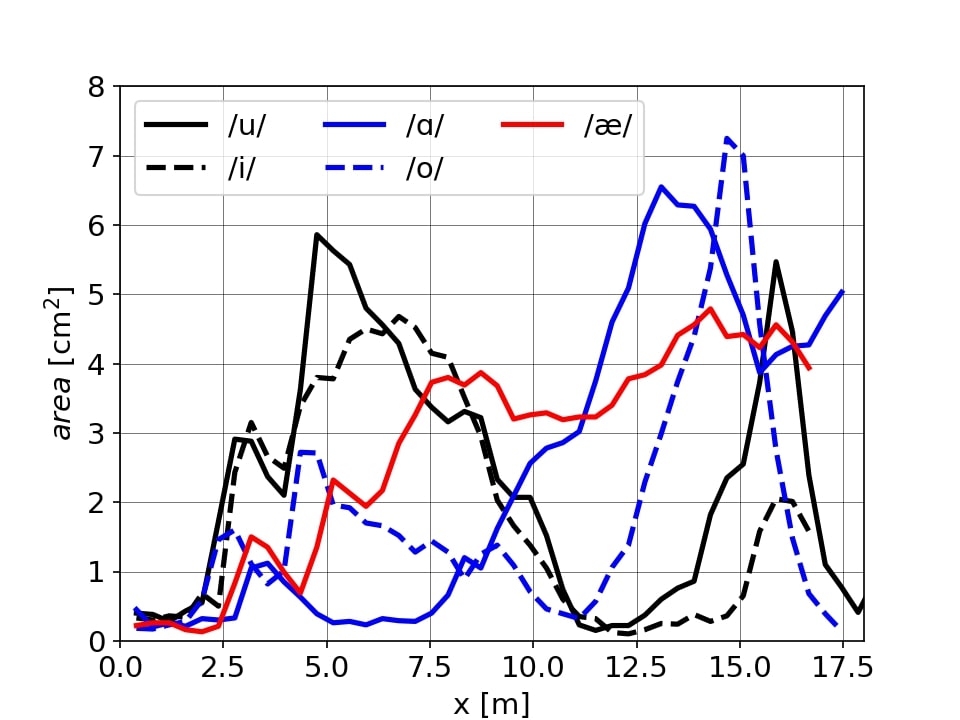} 
    \caption{Surface area function for each geometrical segment of the used vocal tract.}
    \label{fig:trf-vt11}
\end{figure}

The vocal tract was conformally attached to the larynx. The connection is formed by two layers of hexahedral cells, with minor influence on wave propagation. The right edge of the vocal tract was attached to the downstream free field.\deleted{, where at distance 1 cm from the end of the vocal tract the probe MIC1 was placed.} Five geometric models of vocal tracts \added{with different number of segments (42-46)} were created, see Fig.~\ref{fig:trf-vt11}. \deleted{For some vowels, a longer vocal tract is characteristic and thus it is necessary to use a higher number of segments (42-46).} \added{The red arrow in Fig.~\ref{fig:colors} shows how is the interpolation done from the CFD computational domain to the acoustical one.}

The element length $\Delta l^{a}$ of the acoustic mesh and time step $\Delta t^{a}$ for the aeroacoustic simulation are given by estimations \citep{Hueppe-DISS2012} $\Delta l^{a} \le \frac{c_0}{20 f_{{\rm max}}}=3.43~{\rm mm},~~\Delta t^{a} \le \frac{1}{20 f_{{\rm max}}}=1\cdot 10^{-5}~ {\rm s}$ assuming that 20 linear finite elements per one acoustic wavelength are sufficient. In this case, the spatial discretization is limited by $3.43~ \rm mm$ and time step by $1\cdot 10^{-5}~\rm s$ to resolve properly acoustic frequencies up to $f_{{\rm max}}=5~\rm kHz$. If the condition is not satisfied, then the acoustic results are affected by high dissipation and dispersion \citep{Kaltenbacher-book2018}. \deleted{Hexahedral first order finite elements in the range 11-18k were used.}

The partial differential equation \eqref{050520-2018} for the acoustic potential $\psi^{a}$, which will be solved numerically in the acoustic domain, is equipped with zero initial conditions and boundary condition

\begin{equation}
    \nabla \psi^{a} \cdot \mathbf{n} = 0,
\end{equation}

\noindent where $\mathbf{n}$ is the outward unit normal. The boundary condition can be interpreted as a perfect reflection of a sound wave from a barrier, also the condition is called "sound hard". In these computations, the sound hard condition is applied at all solid boundaries except the inflow and outflow, where PML is used.

  \subsection{Numerical solution}
The numerical solution of the aeroacoustic problem proceeds in the following steps \citep{Schoder-JTCA2019}:

  \begin{itemize}
    \item The unsteady flow field in the larynx is computed in OpenFOAM on a fine CFD mesh over 20 periods of vocal fold oscillation.

\item The aeroacoustic sources in the larynx are computed by openCFS\footnote{https://gitlab.com/openCFS} \citep{Schoder-DISS2018,Schoder-ARXIV2022b}, and conservatively interpolated on the coarse CAA mesh \citep{Schoder-JTCA2021,Schoder-AIAA2019}.

\item In the last step, the wave propagation is simulated by openCFS on the coarse CAA mesh \citep{Schoder-JASA2020}.

\end{itemize}


The computational time needed for one CAA simulation is much lower than for one CFD simulation, about five hours on a single CPU core compared to 34 days on 20 cores. The conservative interpolation of the sound source from the very fine CFD mesh (with 2.2M elements) to the coarser CAA mesh (with \replaced{11k-18k}{33k} elements) was performed by the cfsdat tool (part of the openCFS).

  \subsection{CAA results}
  
   \subsubsection{Sound sources (time domain)}
   
\begin{figure}[ht]
\centering
    
      \includegraphics[width=1\linewidth]{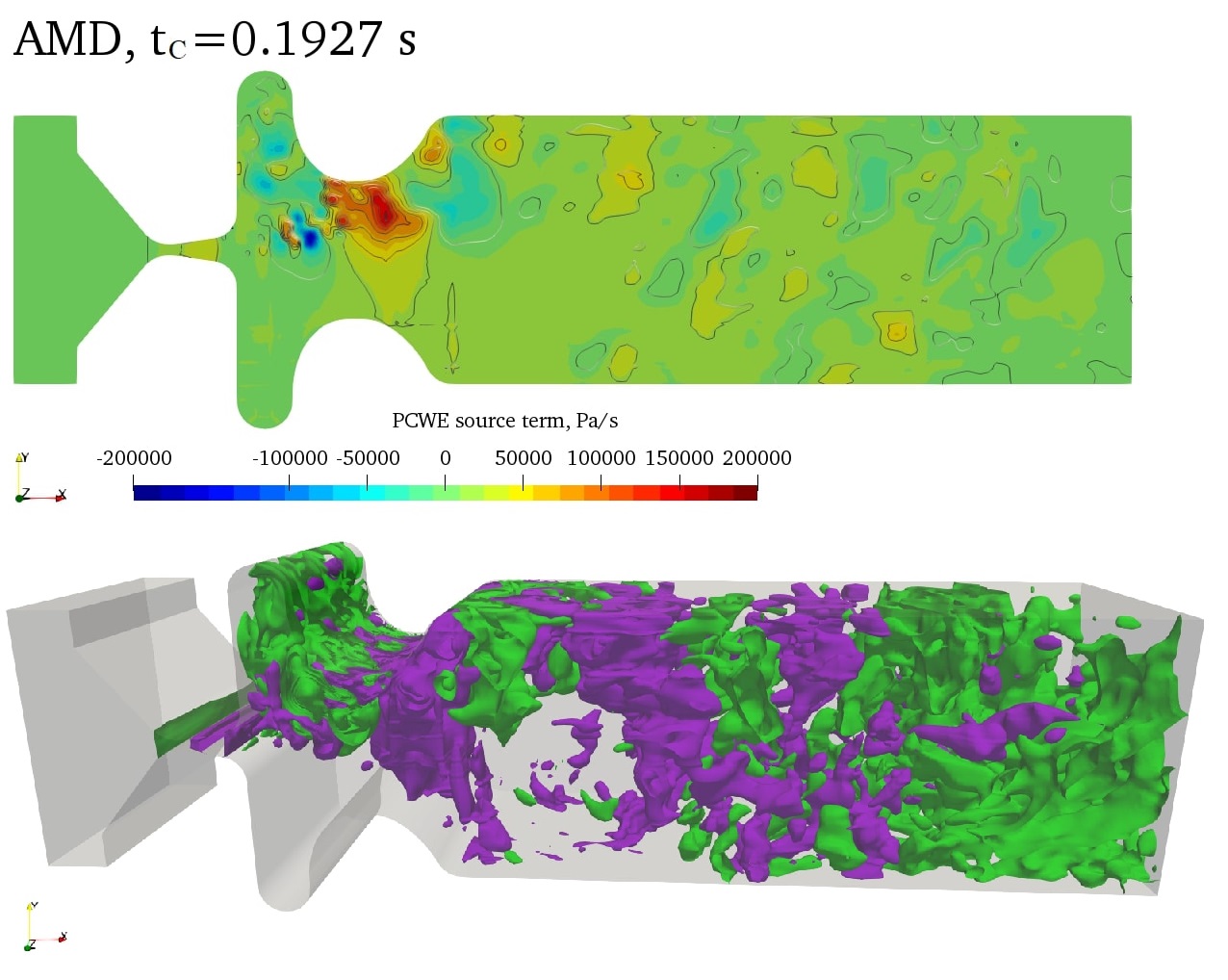} 
   
    \caption{Aeroacoustic source term (${\rm D}p^{ic}/{\rm D}t$) computed by \eqref{050520-2018} at $t_C$; mapped to the domain with closed-divergent vocal folds positions. Twenty iso-surfaces in the range $\pm2\cdot10^{5}{\rm~Pa/s}$ are shown (positive-purple ones, negative-green ones).}
    \label{fig_tt1}
\end{figure}

   The distribution of the aeroacoustic sources in the computational domain covering the larynx varies throughout the vocal fold oscillation period. The jet is surrounded by spots of strong positive and negative acoustic sources related to turbulent eddies created from shear layers of the jet. Fig.~\ref{fig_tt1} shows the aeroacoustic sound source distribution corresponding to the closed-convergent position of vocal folds, which are visualized on the CFD mesh with closed-divergent position of vocal folds. The aeroacoustic sources computed on the moving geometry with oscillating vocal folds are mapped to a fixed geometry. This is done because the current version of the acoustic solver cannot handle moving meshes. 

\begin{figure}[ht]
\centering
      \includegraphics[width=1\linewidth]{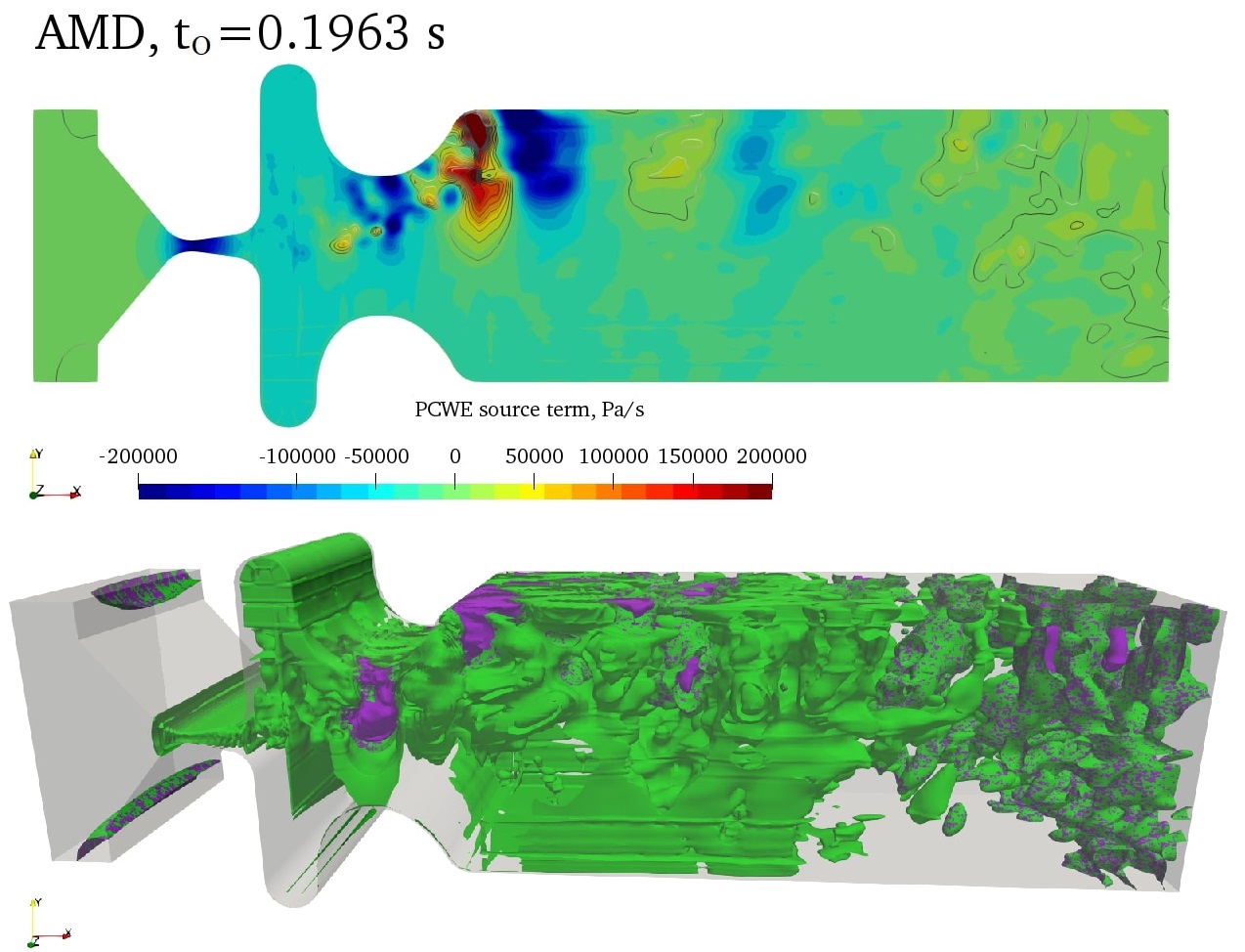} 
   
    \caption{Aeroacoustic source term (${\rm D}p^{ic}/{\rm D}t$) computed by \eqref{050520-2018} at $t_O$; mapped to the domain with closed-divergent vocal folds positions. Twenty iso-surfaces in the range $\pm2\cdot10^{5}{\rm~Pa/s}$ are shown (positive-purple ones, negative-green ones).}
    \label{fig_tt3}
\end{figure}

Fig.~\ref{fig_tt3} shows sound source distribution when vocal folds are fully open, and the results are mapped to the closed-divergent domain. A strong sound source is observed at the glottis, especially on the 2D view, while the 3D view shows preferred directions. This can be caused by the presence of high turbulent viscosity. Compared to adducted vocal folds, the sound sources in the larynx are stronger. 

   \subsubsection{Sound sources (frequency domain)}

\begin{figure}[ht]
\centering
      \includegraphics[width=1\linewidth]{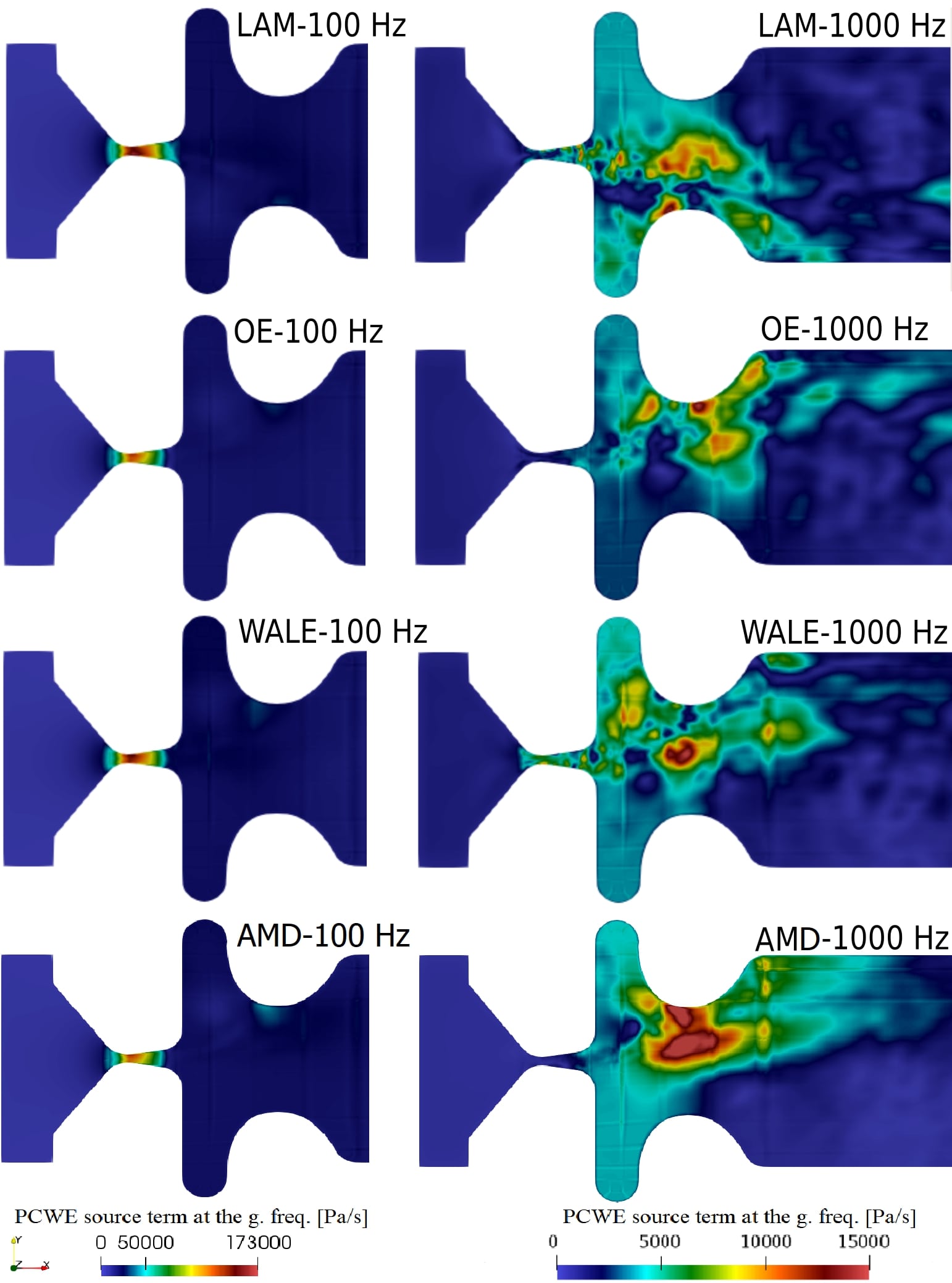}

    \caption{PCWE sound sources in the mid-coronal plane.}
    \label{fig:tt3}
\end{figure}

The conversion from the time to frequency domain was made by the field Fast Fourier Transform (field FFT), which brings insight into the spatial distribution of the aeroacoustic sources at distinct frequencies related somehow to human phonation. The first row in Fig.~\ref{fig:tt3} shows aeroacoustic sources at the fundamental frequency (frequency of vibration of the vocal folds). The strongest sources are located inside the glottis, which is consistent with the theory. Results obtained from the laminar simulation (LAM) show higher intensities than large-eddy simulations (OE, WALE and AMD). This correlates with the flow rate amplitude, which is also higher in the laminar case (see Fig.~\ref{QPUx}). The second row shows the higher harmonic frequency $f_9=1000~ \rm Hz$. Aeroacoustic sources in frequency ranges are firstly observed within the glottis\deleted{in accordance with the theory of}. At these higher frequencies, the dominant aeroacoustic sources do not occur within the glottis but in the places where the fast glottal jet interacts with the ventricular folds and with the slowly moving recirculating air in the supraglottal volume.

\begin{figure}[ht]
 \centering

\includegraphics[width=1\linewidth]{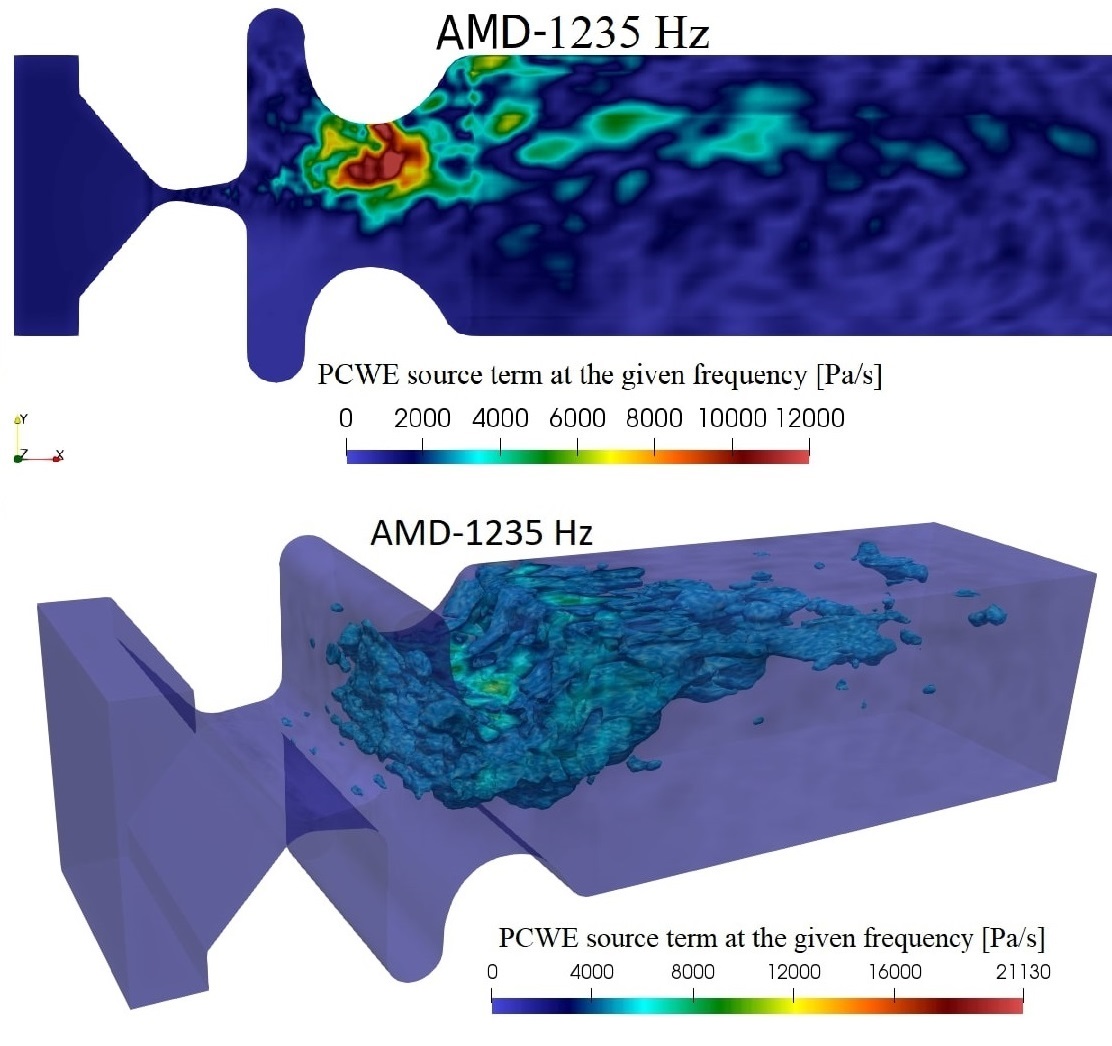} 

    \caption{2D/3D distribution of PCWE sound sources at $f$ = 1235 Hz.}
    \label{fig:as_all}
\end{figure}

Fig.~\ref{fig:as_all} shows the 2D/3D spatial distribution of sound sources in the supraglottal volume at $f=1235~{\rm Hz}$. At the non-harmonic frequency, the acoustic sources are distributed further downstream. The integrity of sound sources can be observed, along with several local sound spots at the superior (trailing) edge of vocal folds.

The visualization of sound sources based on AMD is displayed separately in Fig.~\ref{fig:as_all}. The sound source distribution at $f=100~{\rm Hz}$ in the case based on the AMD subgrid-scale model is most similar to the simulation with OE, and the intensity of the sound sources at $f=100~{\rm Hz}$ is also lower than with the LAM model. On the other side, the intensity of sound sources at $f=1000~{\rm Hz}$ is 2.5-4x higher compared to the LAM, OE and WALE cases. The explanation may be hidden in the presence of higher turbulence intensity corresponding to the fully open glottis. The 2D visualization of the sound source distribution at $f=1235~{\rm Hz}$ shows five times weaker sound sources within ventricles than at the higher harmonic frequency $f=1000~{\rm Hz}$.

\subsubsection{Wave propagation (time domain)}

The wave propagation was studied based on 20 CAA simulations (five vocal tracts shaped by the given vowel and four CFD simulations). In the following, the acoustic pressure fields $p^{a}(\mathbf{x},t)$ will be analyzed as a solution of equation \eqref{050520-2018} and \eqref{040520-1406}. 

\begin{figure}[ht]
 \centering

\includegraphics[width=1\linewidth]{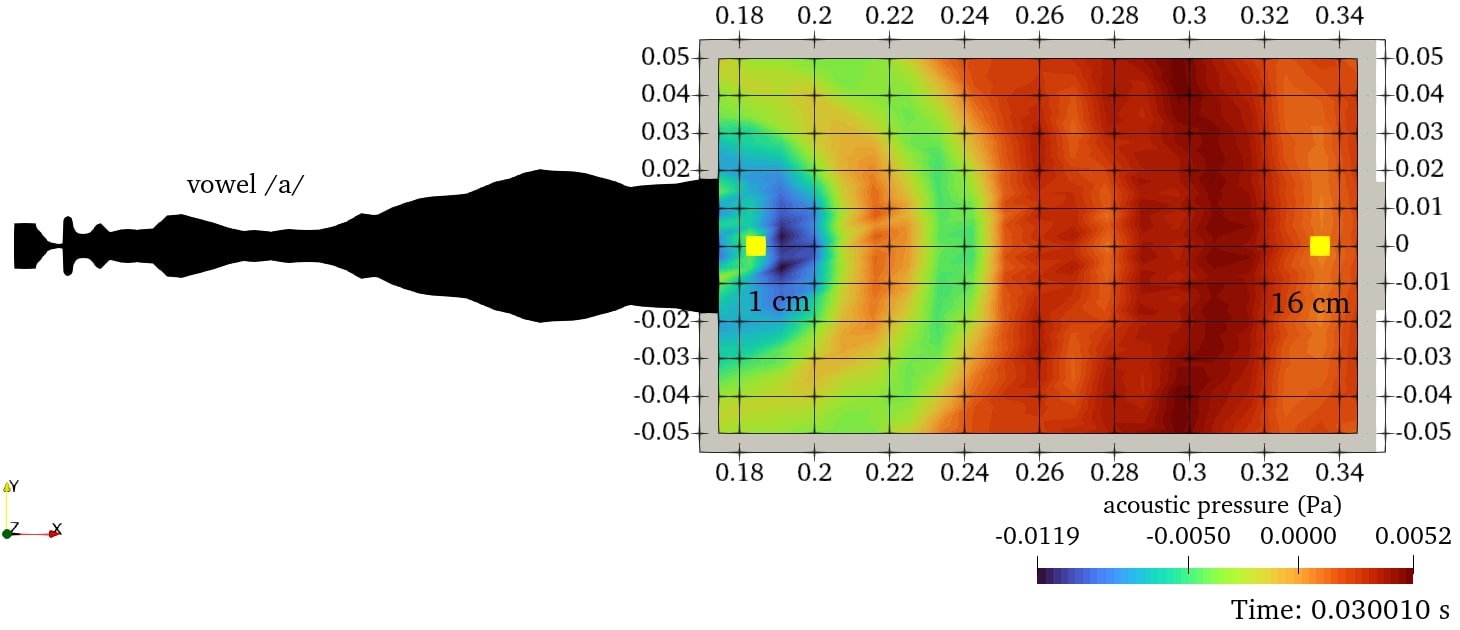} 
    \caption{Acoustic wave propagation in radiation (free) field after 3 periods of vocal folds oscillation (complete case includes 20 periods).}
    \label{fig:wp-waves}
\end{figure}

The acoustic pressures have been \replaced{recorded}{tracked} 1 and 16 cm from lips \added{to create audio recordings. Fig.~\ref{fig:wp-waves} also demonstrates that the two PML layers damped the acoustic waves on walls properly since no reflections are seen.}

Tab.~\ref{tab:prmsSPL} compares individual cases in terms of $p_{\mathrm{rms}}^{a}=\sqrt{\frac{1}{T} \int (p^{a})^2 {\rm d}t}$ and $SPL=20 \log_{10} \bigg ( \frac{p_{{\rm rms}}^{a}}{p_{{\rm ref}}^{a}} \bigg )$, where $p_{{\rm ref}}^{a}=20~\rm \mu Pa$ is the hearing threshold. The SPL value of the radiated sound can be used as a measure of the acoustic energy transferred from the larynx through the vocal tract, and using this value the impact of the subgrid-scale model on aeroacoustics can be analyzed. Minor differences in SPL are observed between the \replaced{front open}{back close} vowel /u/ and \replaced{back close}{front open} vowel /\textipa{A}/. The simulations of front open/mid vowel /æ/ transferred most energy of all simulated vowels. The simulations based on the AMD model predicted the highest SPL of all vowels, except the back-close /u/ and front-close /i/. This may lead to the conclusion that WALE model can transfer more energy in simulations with the so-called close vowels.

\begin{table}[ht]
\caption{Root-mean-square acoustic pressures [Pa] and sound pressure levels [dB] at 16 cm for all simulated vowels and SGS models}
\vspace{0.65cm}
    \centering
    \begin{tabular}{c|c|c|c | c | c | c | c | c | c}
    
Case & & $p_{\mathrm{rms}}^{a}$ & $SPL$ & & $p_{\mathrm{rms}}^{a}$ & $SPL$ & & $p_{\mathrm{rms}}^{a}$ & $SPL$  \\
\hline
LAM & u & 0.0019 & 39.69 & i & 0.0026 & 42.12 & æ & 0.0052 & 48.37 \\
OE & u & 0.0013 & 36.58 & i & 0.0011 & 35.16 & æ & 0.0021 & 40.45 \\
WALE & u & 0.0017 & 38.43 & i & 0.0022 & 40.68 & æ & 0.0049 & 47.84 \\
AMD & u & 0.0018 & 39.22 & i & 0.0016 & 38.01 & æ & 0.0051 & 48.12 \\

\hline
LAM & \textipa{A} & 0.0031 & 43.69 & o & 0.0019 & 39.48   \\
OE & \textipa{A} & 0.0016 & 37.89  & o & 0.0016 & 38.31  \\
WALE & \textipa{A} & 0.0028 & 42.80 & o & 0.0018 & 38.89   \\
AMD & \textipa{A} & 0.0032 & 44.13 & o & 0.0026 & 42.23   \\

    \end{tabular}
    \label{tab:prmsSPL}
\end{table}
   
   \subsubsection{Wave propagation (frequency domain)}

This section deals with frequency spectra computed at the \replaced{1}{16} cm distance from the mouth. This kind of analysis can highlight fundamental, harmonic and non-harmonic frequencies, accompanied by broadband noise. The FFT analyses were performed on the signal spanning over 20 periods of the vocal fold oscillation (1 period = 10~ms = 1000~samples), and thus the frequency resolution is $\Delta f = 5~ \rm Hz$. 

\begin{figure}[ht]
 \centering

\includegraphics[width=1\linewidth]{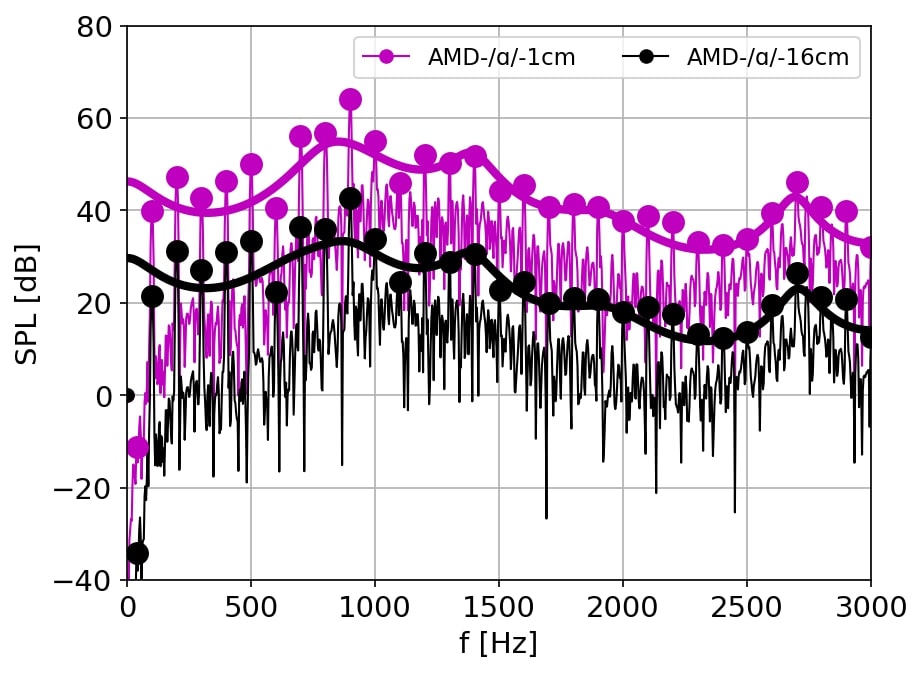} 
    \caption{Acoustic sound spectra from the simulation of vocalization of /\textipa{A}/ in 1 cm and 16 cm from lips after 20 periods of vocal folds oscillation.}
    \label{fig:wp-waves2}
\end{figure}

\added{Fig.~\ref{fig:wp-waves2} confirms that the signal obtained by the probe at 1~cm is not affected by lips as one could expect. These models of phonation prescribing 350 Pa at lungs can be considered to be quite phonation, in other case it would be recommended to use at least 3~cm. The sound pressure levels on all components are decreased by 19 dB moving the probe at 16 cm from lips. The envelope overlaying black peaks corresponds to the linear predictive coding (LPC) curve \citep{Pavlidi-IEEE2013} widely used to provide a smooth spectral envelope of audio and voice recordings in purpose to find positions of formants.}

\begin{figure}[ht]
    \centering
      \includegraphics[width=1\linewidth]{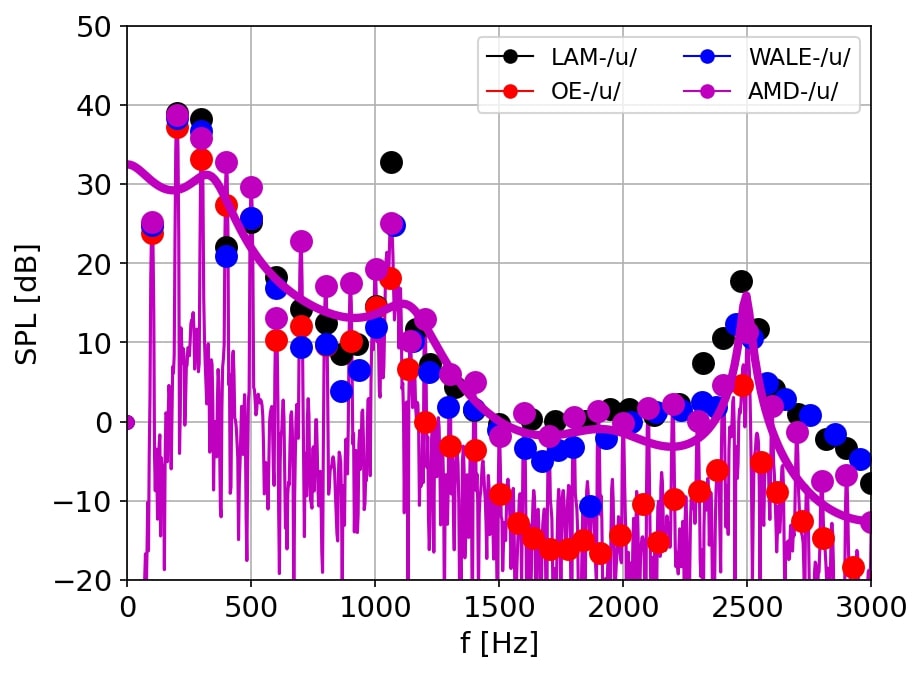}
    \caption{Acoustic sound spectra from the simulation of vocalization of /u/ at 16 cm from lips.}
    \label{mic1u}
\end{figure}

The \added{acoustic} spectra will be analyzed vowel by vowel \added{showing differences caused by turbulence subgrid-scale models}:

\textit{Vowel} /u/. Fig.~\ref{mic1u} shows aeroacoustic spectrum based on the CFD simulation with\deleted{different subgrid-scale models} \added{three subgrid-scale models and one "laminar" case without SGS model. Note that the laminar simulation should be considered as the least accurate case, which disregards the effect of small-scale turbulence. However, this approach is still often used in the modeling of laryngeal flow and the comparison against more accurate LES simulations is interesting.} SPL at fundamental frequency $f_o=100~{\rm Hz}$ and higher harmonics $f_1=200~{\rm Hz}, f_2=300~{\rm Hz}$ and so forth are well visible. SPL at $f_o$ is lower than at $f_1$ and $f_2$\added{; when the signal is processed immediately behind the vocal folds the SPL at $f_o$ would be definitely the highest.}\deleted{Scientific groups ${\citep{Falk-FP2021,Schoder-JASA2020}}$ report the same trend with the first harmonic $L_{f_1}$ higher than $L_{f_o}$.} The second \added{and third} formant computed by the simulation with AMD has the same SPL compared to the case with WALE.\deleted{At the third formant, on the contrary, WALE is higher by 13\% than AMD. This trend occurs only for vowels /u, \textipa{A}/, even though the vocal tracts for /u, \textipa{A}/ have different shapes (see Fig.~\ref{fig:trf-vt11}).}

\begin{figure}[ht]
    \centering
      \includegraphics[width=1\linewidth]{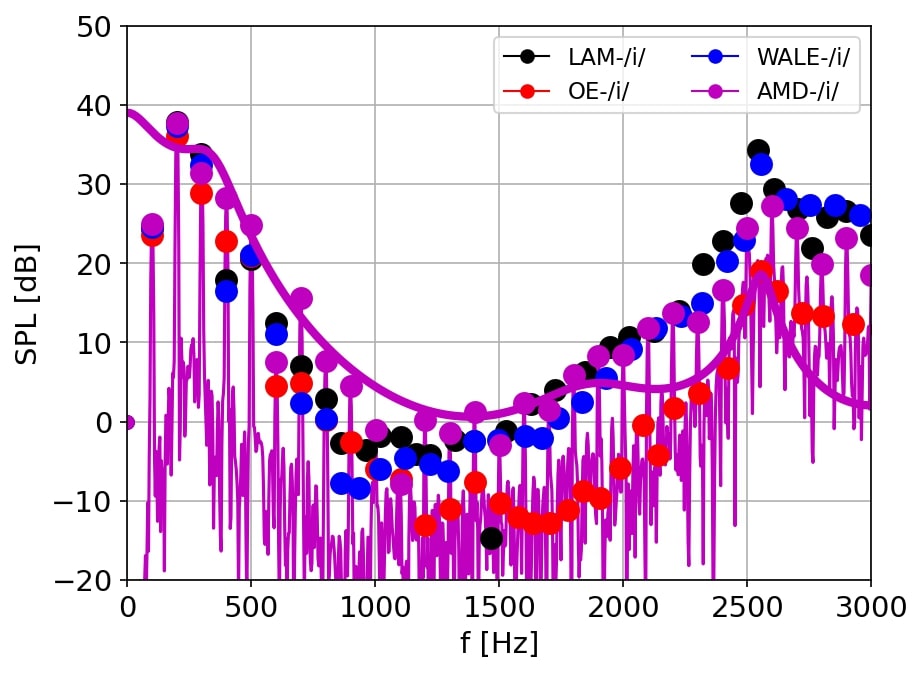} 
    \caption{Acoustic sound spectra from the simulation of vocalization /i/ at 16 cm from lips.}
    \label{mic1i}
\end{figure}

\textit{Vowel} /i/. The second aeroacoustic spectrum is plotted in Fig.~\ref{mic1i}. Simulations with OE\deleted{performed on vocal tracts /u, i/ predict the lowest SPL at the fundamental frequencies} \added{predicted for all vowels the lowest SPL, that can be related to the well-known overpredicting of turbulent viscosity.}

\begin{figure}[ht]
    \centering
      \includegraphics[width=1\linewidth]{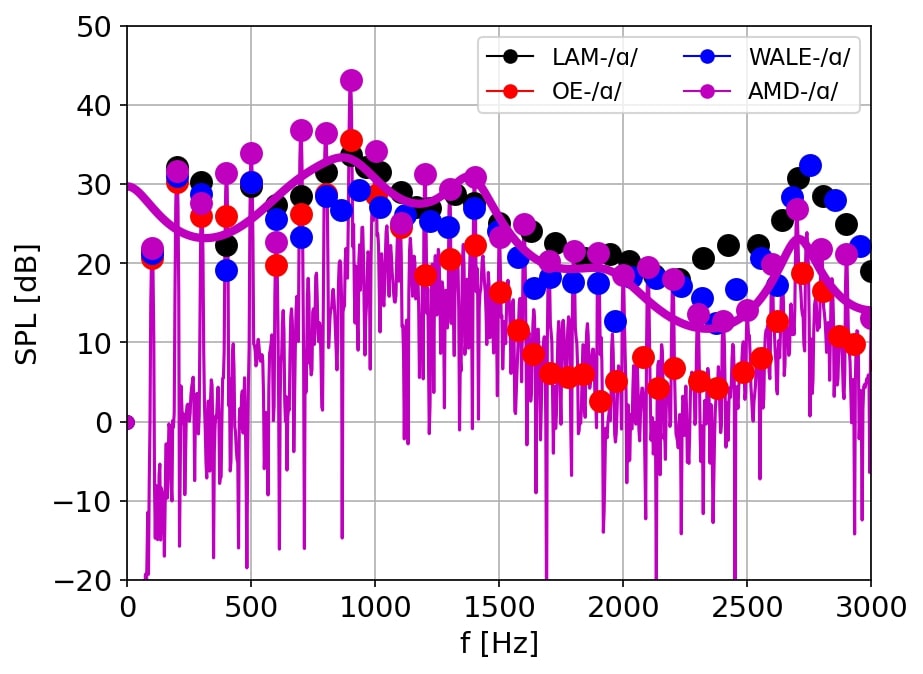}
    \caption{Acoustic sound spectra from the simulation of vocalization of /\textipa{A}/ at 16 cm from lips.}
    \label{mic1a}
\end{figure}

\textit{Vowel} /\textipa{A}/. The spectrum is in Fig.~\ref{mic1a}.\deleted{, where is shown that SPLs at fundamental frequency remain at the same level for all models. This happened only twice, in cases /\textipa{A}, æ/ for open and mid-open vowels, when the tongue is pressed down most. The close distance between formants $F_1-F_2$ is typical for vowels /u, \textipa{A}, o/, but in the simulation of /\textipa{A}/ the second formant around $1300~\rm Hz$ was not detected. However, in the case of AMD, it appears that the second formant may be found. On the other hand, the third formant is clearly visible and presents the same behavior as in /u/, i.e., a 3-10 dB lower value of AMD compared to WALE.} \added{The close distance between formants $F_1-F_2$ is typical for vowels /u, \textipa{A}, o/. Formants are moved a bit to the higher frequency range since the back side of tongue is pressed down most. Open/mid-open vowels /\textipa{A}, æ/ were captured best by our models of phonations and are the most recognizable in the recordings\footnote{\added{See supplementary material at [URL will be inserted by AIP] for listening all the simulated vowels based on different subgrid-scale turbulence models.}}.}

\begin{figure}[ht]
    \centering
      \includegraphics[width=1\linewidth]{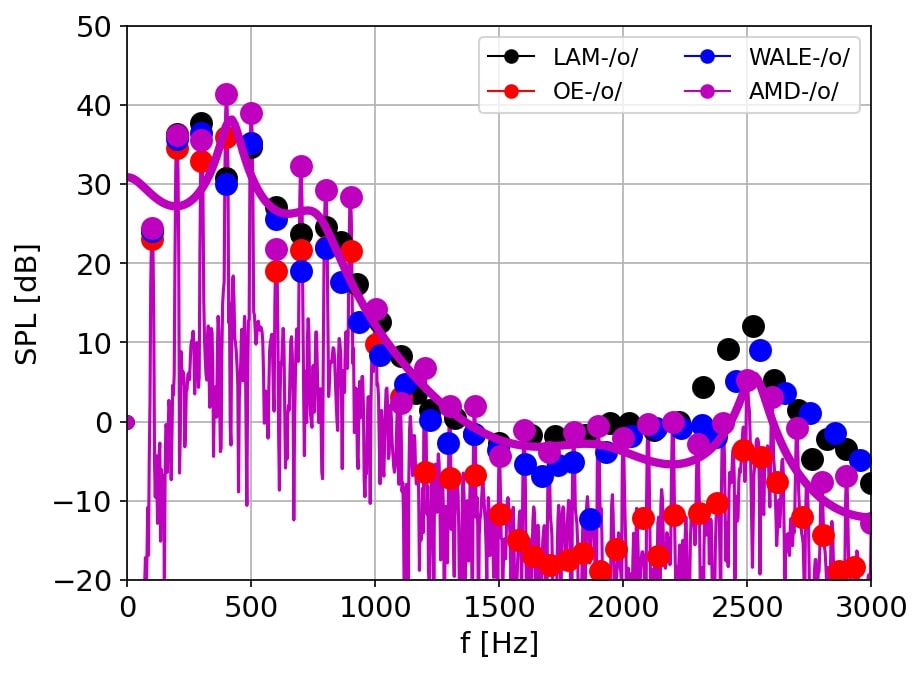} 
    \caption{Acoustic sound spectra from the simulation of vocalization of /o/ at 16 cm from lips.}
    \label{mic1o}
\end{figure}

\textit{Vowel} /o/. Fig.~\ref{mic1o} shows the aeroacoustic spectrum for the mid-close vowel, where is typical that $F_{2}-F_{3}$ are far apart.\deleted{with the widest passage of the throat ($7.25~{\rm cm^2}$) during phonation (see Fig.~\ref{fig:trf-vt11})} \added{Simulations with WALE predicted the third formant of all vowels the most visible. Human ears are sensitive to first two formants at lower frequencies to distinguish vowel. From that reason can be AMD model more valuable for voice research.}

\begin{figure}[ht]
    \centering
      \includegraphics[width=1\linewidth]{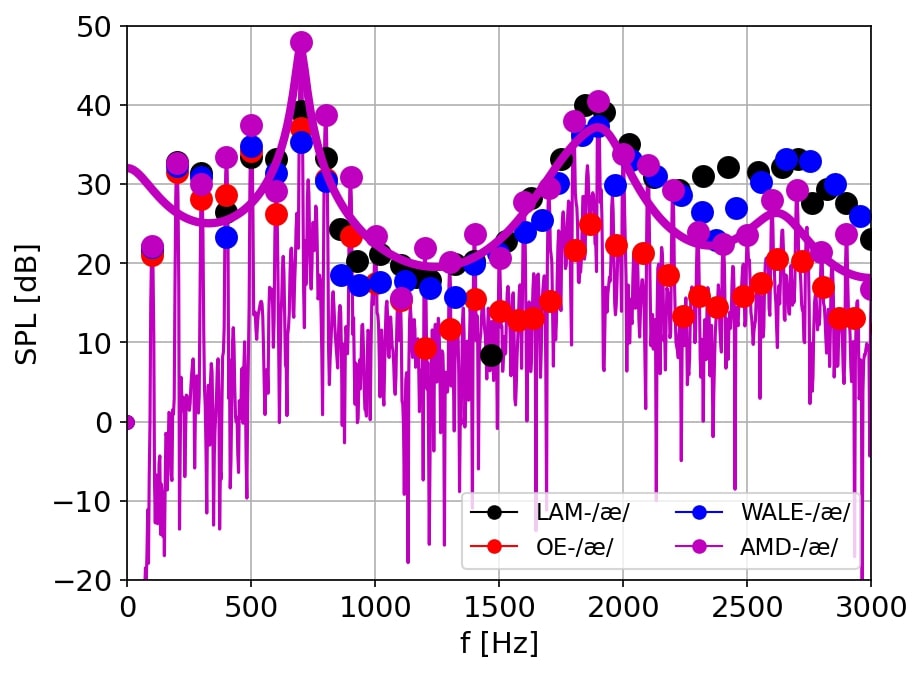}
    \caption{Acoustic sound spectra from the numerical simulation of vocalization of /æ/ at 16 cm from lips.}
    \label{fig:mic1ae}
\end{figure}

\textit{Vowel} /æ/. Fig.~\ref{fig:mic1ae}\deleted{shows the aeroacoustic spectrum, which transfers most acoustic energy, i.e., 71 dB with the AMD model. The first formant predicted by AMD is by 13 dB higher than the formant predicted by the WALE model. The formants in the high-frequency bandwidth are on the same level for AMD and WALE. For all vowels except /æ/, at 600 Hz a significant drop in SPL was detected.} \added{shows the spectrum corresponding to the highest total SPL from simulated vowels, up to 48 dB. There is the next vowel with fairly clear vowel distinctness in recordings, especially for AMD. It should be noted that only the highly located SPLs of first two fromants are not enough for clarity, as it was observed in previous spectra where WALE and AMD had very similar SPLs. Other frequency components somehow contribute to the final tuning of the vowel.}

\begin{figure}[ht]
    \centering
      \includegraphics[width=1\linewidth]{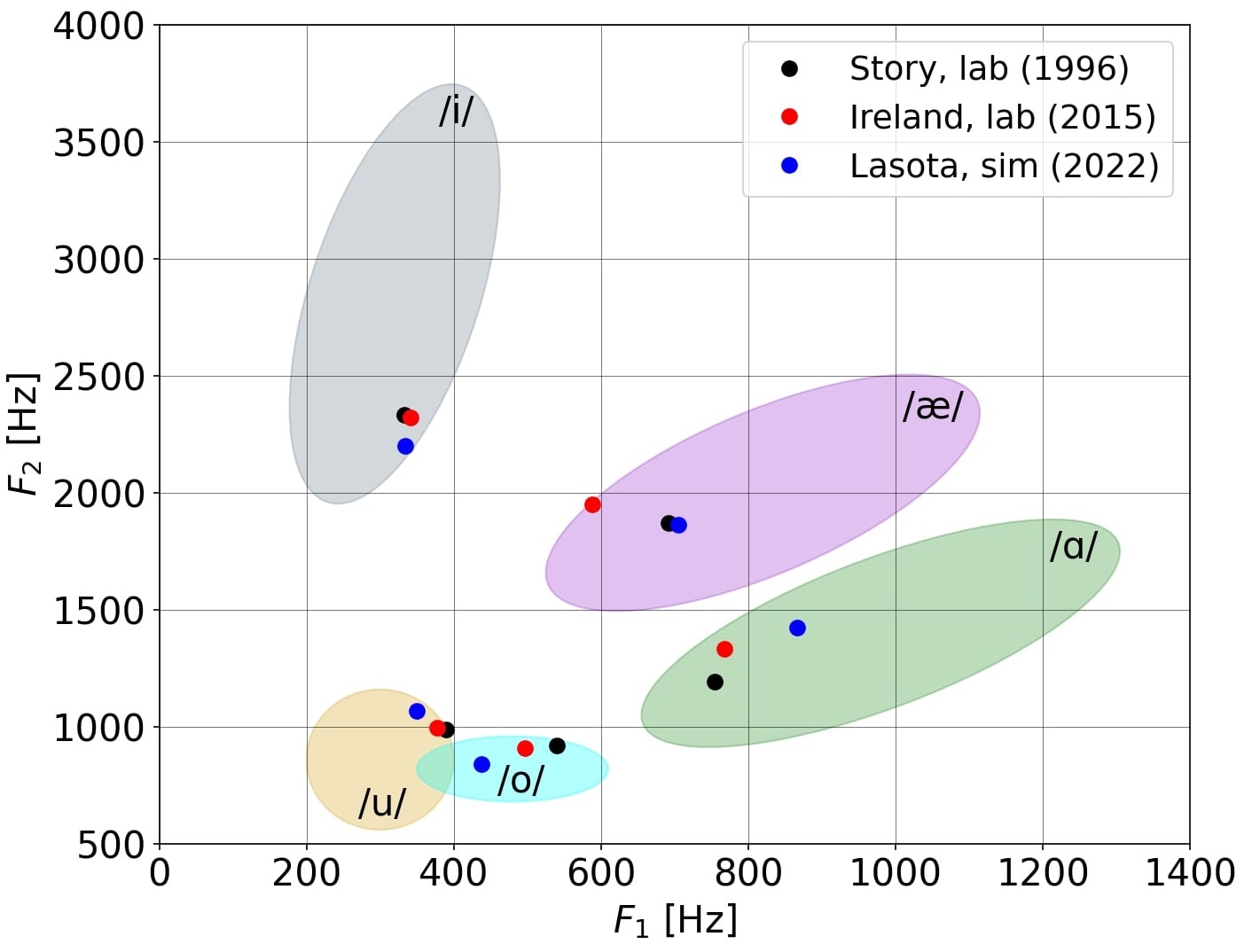}
    \caption{Formant ranges measured by \citet{Peterson-JASA1952}, formant locations obtained by \citet{Story-JASA1996} and \citet{Ireland-FBB2015}; all in comparison with the simulated vowels.}
    \label{fig:finer}
\end{figure}

Fig.~\ref{fig:finer} shows the formant ranges (colorful ellipses) measured by \citet{Peterson-JASA1952}\deleted{together with the currently simulated formants}\added{, Ireland's and Story's formants from measuruments of natural speech \citep{Story-JASA1996, Ireland-FBB2015} and our simulated formants $F_{1}-F_{2}$ computed precisely from local maxima of LPC curves. All simulated vowels lie inside the measured ranges}.\deleted{The simulated vowels /u, æ/ lie inside the measured ranges. Vowels /\textipa{A}, o/ lie in different ellipses and /i/ is outside any measured ranges. The vowel tract geometry could be relatively easily modified to shift the location of formants where we need. The prolonged supraglottal space, which is mounted between the vocal folds and vocal tracts, shifted the first formants just slightly, not more than 5-60~Hz.} Nevertheless, Fig.~\ref{fig:finer} confirms the usability of grids based on the circular vocal tracts \citep{Story-JASA1996}.

\section{Conclusion}

\deleted{The concluding assessment regarding the usage of turbulence subgrid-scale models in numerical modelling of human phonation can be formulated as follows: }\added{In this article, models of phonation based on large-eddy simulations and their subgrid-scale models contributing to the solution of turbulent flow were presented. It was found that} the OE model overpredicts the turbulent viscosity in regions where shear is dominant, i.e., in the boundary layer adjacent to the vocal folds and in the shear layers of the glottal jet (Fig.~\ref{fig:viscosity_all0}). The difference in glottal flow rate among the simulations is clearly induced by the subgrid-scale model, which adds the turbulent viscosity to the molecular viscosity of air and hinders the airflow in the glottis (Fig.~\ref{QPUx}). The WALE model produced zero eddy viscosity in cases of pure shear flow (Fig.~\ref{fig:nutxz1}), and hence the flow simulation with WALE predicted by 5\% higher maximum transglottal flow rate than AMD. Despite this fact, the phonation simulation based on the AMD model transferred more \added{or equal }energy in terms of total sound pressure level than WALE for all vowels except the front-close vowel /i/ (Tab.~\ref{tab:prmsSPL}). 

The WALE model, which is known to handle turbulent viscosity at the near-wall and high-shear regions more precisely than the OE model (Figs.~\ref{fig:viscosity_all0}-\ref{fig:nutxz1}), resulted in higher \replaced{SPLs}{total sound pressure levels} than OE in all cases\deleted{except the close-back vowel /u/} (Tab.~\ref{tab:prmsSPL}). The OE model gives acceptable results in general, but peaks of frequency formants are hardly visible and weaker compared to the WALE or AMD model (Figs.~\ref{mic1u}-\ref{fig:mic1ae}). The WALE model amplified third formants in high-frequency bandwidth most of all the \added{subgrid-scale} models (Figs.~\ref{mic1u}-\ref{fig:mic1ae}). However, the third formant is not crucial for vowel characterization. 

\deleted{The AMD model seems to be a very promising successor to the WALE model in modelling laryngeal flow, since the AMD model resulted in significantly higher harmonic frequencies up to the second formant for all studied cardinal vowels (Figs.~\ref{mic1u}-\ref{fig:mic1ae}). This finding could be related to predicted features of the AMD model: consistency with the exact subgrid-scale stress tensor $\tau_{ij}$, no requirements on the approximation of the LES filter width $\Delta$ and usability on an anisotropic mesh.}

\added{On the other side, the AMD model resulted in slightly higher sound pressure levels at harmonic and formant frequencies up to the second formant for all vowels (Figs.~\ref{mic1u}-\ref{fig:mic1ae}). First formants are even amplified in cases /\textipa{A}, æ/ at least by 8 dB (Figs.~\ref{mic1a}, \ref{fig:mic1ae}), which surely contributed to higher quality of simulated vowels. Recordings are enclosed to this article to let readers evaluate themselves the clarity of simulated vowels. Please note that signals were recorded from 1 cm from lips for comfortable volume and repeated twenty times in one sample. 1 cm from lips is probably on the edge of applicability having clear wavefronts (Fig.~\ref{fig:wp-waves}). Based on these recordings of acoustic pressures, it can be concluded that the model of phonation employing AMD is much closer to natural speech than other models (WALE, OE, and LAM). 

The AMD model seems to be a very promising successor to the WALE model, which is the most common choice in turbulence modelling of laryngeal flow. This positive finding can be attributed to beneficial features of the AMD model: consistency with the exact subgrid-scale stress tensor $\tau_{ij}$, no requirements on the approximation of the LES filter width\deleted{$\Delta$} and usability on anisotropic meshes.}

\begin{acknowledgments}
The research was supported by the Student Grant Scheme at the Technical University of Liberec through project no. SGS-2022-3016. The Graz group acknowledges support from the ÖAW research grant "Understanding voice disorders", received from "Dr. Anton Oelzelt-Newin'sche Stiftung".
\end{acknowledgments}






\bibliography{Manuscript.bib}

\begin{thebibliography}{59}
\def\enquote#1{``#1,''}
\def\plainquote#1{``#1''}
\expandafter\ifx\csname natexlab\endcsname\relax\def\natexlab#1{#1}\fi
\providecommand{\dourl}[1]{\href{http://#1}{\nolinkurl{#1}}}
\providecommand{\bibinfo}[2]{#2}
\providecommand{\noopsort}[1]{}
\providecommand{\switchargs}[2]{#2#1}
  \def\eatspace #1{#1}

\bibitem[{Abkar and Moin(2017)}]{Abkar-2017BLM}
\bibinfo{author}{Abkar, M.},  and \bibinfo{author}{Moin, P.}
  (\textbf{\bibinfo{year}{2017}}). \enquote{\bibinfo{title}{Large-eddy
  simulation of thermally stratified atmospheric boundary-layer flow using a
  minimum dissipation model}} \bibinfo{journal}{{B}oundary-{L}ayer
  {M}eteorology} \textbf{165}(3), \bibinfo{pages}{405--419},
  \dodoi{10.1007/s10546-017-0288-4}.

\bibitem[{Agarwal \emph{et~al.}(2003)Agarwal, Scherer, and
  Hollien}]{Agarwal-JV2003}
\bibinfo{author}{Agarwal, M.}, \bibinfo{author}{Scherer, R.},  and
  \bibinfo{author}{Hollien, H.} (\textbf{\bibinfo{year}{2003}}).
  \enquote{\bibinfo{title}{The false vocal folds: shape and size in frontal
  view during phonation based on laminagraphic tracings}}
  \bibinfo{journal}{Journal of Voice} \textbf{17}(2), \bibinfo{pages}{97--113},
  \dodoi{10.1016/S0892-1997(03)00012-2}.

\bibitem[{Anghel and Iacobescu(2013)}]{Anghel-ICM2013}
\bibinfo{author}{Anghel, M.-A.},  and \bibinfo{author}{Iacobescu, F.}
  (\textbf{\bibinfo{year}{2013}}). \enquote{\bibinfo{title}{The influence of
  temperature and co2 in exhaled breath}} in \emph{\bibinfo{booktitle}{16th
  International Congress of Metrology}}, \bibinfo{organization}{EDP Sciences},
  p. \bibinfo{pages}{10012}.

\bibitem[{Avhad \emph{et~al.}(2022)Avhad, Li, Wilson, Sayce, Chang, Rousseau,
  and Luo}]{Avhad-F2022}
\bibinfo{author}{Avhad, A.}, \bibinfo{author}{Li, Z.}, \bibinfo{author}{Wilson,
  A.}, \bibinfo{author}{Sayce, L.}, \bibinfo{author}{Chang, S.},
  \bibinfo{author}{Rousseau, B.},  and \bibinfo{author}{Luo, H.}
  (\textbf{\bibinfo{year}{2022}}). \enquote{\bibinfo{title}{Subject-specific
  computational fluid-structure interaction modeling of rabbit vocal fold
  vibration}} \bibinfo{journal}{Fluids} \textbf{7}(3),
  \dodoi{10.3390/fluids7030097}.

\bibitem[{Berenger(1994)}]{Berenger-JCP1994}
\bibinfo{author}{Berenger, J.-P.} (\textbf{\bibinfo{year}{1994}}).
  \enquote{\bibinfo{title}{A perfectly matched layer for the absorption of
  electromagnetic waves}} \bibinfo{journal}{Journal of {C}omputational
  {P}hysics} \textbf{114}(2), \bibinfo{pages}{185--200},
  \dodoi{10.1006/jcph.1994.1159}.

\bibitem[{Bodaghi \emph{et~al.}(2021)Bodaghi, Jiang, Xue, and
  Zheng}]{Bodaghi-JBE2021}
\bibinfo{author}{Bodaghi, D.}, \bibinfo{author}{Jiang, W.},
  \bibinfo{author}{Xue, Q.},  and \bibinfo{author}{Zheng, X.}
  (\textbf{\bibinfo{year}{2021}}). \enquote{\bibinfo{title}{Effect of
  supraglottal acoustics on fluid-structure interaction during human voice
  production}} \bibinfo{journal}{Journal of Biomechanical Engineering}
  \textbf{143}(4), \dodoi{10.1115/1.4049497}.

\bibitem[{D{\"o}llinger \emph{et~al.}(2011)D{\"o}llinger, Kobler, A~Berry,
  D~Mehta, Luegmair, and Bohr}]{Doellinger-CB2011}
\bibinfo{author}{D{\"o}llinger, M.}, \bibinfo{author}{Kobler, J.},
  \bibinfo{author}{A~Berry, D.}, \bibinfo{author}{D~Mehta, D.},
  \bibinfo{author}{Luegmair, G.},  and \bibinfo{author}{Bohr, C.}
  (\textbf{\bibinfo{year}{2011}}). \enquote{\bibinfo{title}{Experiments on
  analysing voice production: Excised (human, animal) and in vivo (animal)
  approaches}} \bibinfo{journal}{Current {B}ioinformatics} \textbf{6}(3),
  \bibinfo{pages}{286--304}, \dodoi{10.2174/157489311796904673}.

\bibitem[{Erath and Plesniak(2010)}]{Erath-EF2010}
\bibinfo{author}{Erath, B.},  and \bibinfo{author}{Plesniak, M.}
  (\textbf{\bibinfo{year}{2010}}). \enquote{\bibinfo{title}{An investigation of
  asymmetric flow features in a scaled-up driven model of the human vocal
  folds}} \bibinfo{journal}{Experiments in Fluids} \textbf{49}(1),
  \bibinfo{pages}{131--146}, \dodoi{10.1007/s00348-009-0809-0}.

\bibitem[{Falk \emph{et~al.}(2021)Falk, Kniesburges, Schoder, Jakuba{\ss},
  Maurerlehner, Echternach, Kaltenbacher, and D{\"o}llinger}]{Falk-FP2021}
\bibinfo{author}{Falk, S.}, \bibinfo{author}{Kniesburges, S.},
  \bibinfo{author}{Schoder, S.}, \bibinfo{author}{Jakuba{\ss}, B.},
  \bibinfo{author}{Maurerlehner, P.}, \bibinfo{author}{Echternach, M.},
  \bibinfo{author}{Kaltenbacher, M.},  and \bibinfo{author}{D{\"o}llinger, M.}
  (\textbf{\bibinfo{year}{2021}}). \enquote{\bibinfo{title}{3d-fv-fe
  aeroacoustic larynx model for investigation of functional based voice
  disorders}} \bibinfo{journal}{Frontiers in {P}hysiology} \textbf{12},
  \bibinfo{pages}{226}, \dodoi{10.3389/fphys.2021.616985}.

\bibitem[{Ferziger(1998)}]{Ferziger-NMFM1998}
\bibinfo{author}{Ferziger, H.} (\textbf{\bibinfo{year}{1998}}).
  \enquote{\bibinfo{title}{Direct and large eddy simulation of turbulence}}
  \bibinfo{journal}{Numerical {M}ethods in {F}luid {M}echanics} \textbf{16},
  \bibinfo{pages}{53--73}, \dodoi{10.1299/kikaib.66.651_2754}.

\bibitem[{Ffowcs~Williams and Hawkings(1969)}]{Ffowcs-PTRSL1969}
\bibinfo{author}{Ffowcs~Williams, J.~E.},  and \bibinfo{author}{Hawkings,
  D.~L.} (\textbf{\bibinfo{year}{1969}}). \enquote{\bibinfo{title}{Sound
  generation by turbulence and surfaces in arbitrary motion}}
  \bibinfo{journal}{Philosophical Transactions of the Royal Society of London.
  Series A, Mathematical and Physical Sciences} \textbf{264}(1151),
  \bibinfo{pages}{321--342}, \dodoi{10.1098/rsta.1969.0031}.

\bibitem[{Fletcher(1991)}]{Fletcher-FD1991}
\bibinfo{author}{Fletcher, C.~A.} (\textbf{\bibinfo{year}{1991}}).
  \enquote{\bibinfo{title}{Fluid dynamics: The governing equations}} in
  \emph{\bibinfo{booktitle}{Computational Techniques for Fluid Dynamics 2}}
  (\bibinfo{publisher}{Springer: Berlin/Heidelberg}), pp.
  \bibinfo{pages}{1--46}, \dodoi{10.1007/978-3-642-58239-4_1}.

\bibitem[{Georgiadis \emph{et~al.}(2010)Georgiadis, Rizzetta, and
  Fureby}]{Georgiadis-AIAAJ2010}
\bibinfo{author}{Georgiadis, N.~J.}, \bibinfo{author}{Rizzetta, D.~P.},  and
  \bibinfo{author}{Fureby, C.} (\textbf{\bibinfo{year}{2010}}).
  \enquote{\bibinfo{title}{Large-eddy simulation: current capabilities,
  recommended practices, and future research}} \bibinfo{journal}{AIAA journal}
  \textbf{48}(8), \bibinfo{pages}{1772--1784}, \dodoi{10.2514/1.j050232}.

\bibitem[{H{\"u}ppe(2012)}]{Hueppe-DISS2012}
\bibinfo{author}{H{\"u}ppe, A.} (\textbf{\bibinfo{year}{2012}}).
  \enquote{\bibinfo{title}{Spectral finite elements for acoustic field
  computation}} Ph.D. thesis, \bibinfo{school}{Alps-Adriatic University of
  Klagenfurt}.

\bibitem[{H{\"u}ppe \emph{et~al.}(2014)H{\"u}ppe, Grabinger, Kaltenbacher,
  Reppenhagen, Dutzler, and K{\"u}hnel}]{Hueppe-AIAA2014}
\bibinfo{author}{H{\"u}ppe, A.}, \bibinfo{author}{Grabinger, J.},
  \bibinfo{author}{Kaltenbacher, M.}, \bibinfo{author}{Reppenhagen, A.},
  \bibinfo{author}{Dutzler, G.},  and \bibinfo{author}{K{\"u}hnel, W.}
  (\textbf{\bibinfo{year}{2014}}). \enquote{\bibinfo{title}{A non-conforming
  finite element method for computational aeroacoustics in rotating systems}}
  in \emph{\bibinfo{booktitle}{20th AIAA/CEAS Aeroacoustics Conference}}, p.
  \bibinfo{pages}{2739}, \dodoi{10.2514/6.2014-2739}.

\bibitem[{Ireland \emph{et~al.}(2015)Ireland, Knuepffer, and
  McBride}]{Ireland-FBB2015}
\bibinfo{author}{Ireland, D.}, \bibinfo{author}{Knuepffer, C.},  and
  \bibinfo{author}{McBride, S.~J.} (\textbf{\bibinfo{year}{2015}}).
  \enquote{\bibinfo{title}{Adaptive multi-rate compression effects on vowel
  analysis}} \bibinfo{journal}{Frontiers in {B}ioengineering and
  {B}iotechnology} \textbf{3}, \bibinfo{pages}{118},
  \dodoi{10.3389/fbioe.2015.00118}.

\bibitem[{Jasak(1996)}]{Jasak-DISS1996}
\bibinfo{author}{Jasak, H.} (\textbf{\bibinfo{year}{1996}}).
  \enquote{\bibinfo{title}{Error analysis and estimation for the finite volume
  method with applications to fluid flows.}} Ph.D. thesis,
  \bibinfo{school}{Imperial College London}.

\bibitem[{Jiang and Lai(2016)}]{Jiang-book2016}
\bibinfo{author}{Jiang, X.},  and \bibinfo{author}{Lai, C.-H.}
  (\textbf{\bibinfo{year}{2016}}). \emph{\bibinfo{title}{Numerical techniques
  for direct and large-eddy simulations}} (\bibinfo{publisher}{CRC press}).

\bibitem[{Kaltenbacher \emph{et~al.}(2013)Kaltenbacher, Kaltenbacher, and
  Sim}]{Kaltenbacher-JCP2013}
\bibinfo{author}{Kaltenbacher, B.}, \bibinfo{author}{Kaltenbacher, M.},  and
  \bibinfo{author}{Sim, I.} (\textbf{\bibinfo{year}{2013}}).
  \enquote{\bibinfo{title}{A modified and stable version of a perfectly matched
  layer technique for the 3-d second order wave equation in time domain with an
  application to aeroacoustics}} \bibinfo{journal}{Journal of {C}omputational
  {P}hysics} \textbf{235}, \bibinfo{pages}{407--422},
  \dodoi{10.1016/j.jcp.2012.10.016}.

\bibitem[{Kaltenbacher(2018)}]{Kaltenbacher-book2018}
\bibinfo{author}{Kaltenbacher, M.} (\textbf{\bibinfo{year}{2018}}).
  \emph{\bibinfo{title}{Computational Acoustics}}
  (\bibinfo{publisher}{Springer}).

\bibitem[{Kniesburges \emph{et~al.}(2011)Kniesburges, L~Thomson, Barney, Triep,
  {\v{S}}idlof, Hor{\'a}{\v{c}}ek, Br{\"u}cker, and
  Becker}]{Kniesburges-CB2011}
\bibinfo{author}{Kniesburges, S.}, \bibinfo{author}{L~Thomson, S.},
  \bibinfo{author}{Barney, A.}, \bibinfo{author}{Triep, M.},
  \bibinfo{author}{{\v{S}}idlof, P.}, \bibinfo{author}{Hor{\'a}{\v{c}}ek, J.},
  \bibinfo{author}{Br{\"u}cker, C.},  and \bibinfo{author}{Becker, S.}
  (\textbf{\bibinfo{year}{2011}}). \enquote{\bibinfo{title}{In vitro
  experimental investigation of voice production}} \bibinfo{journal}{Current
  {B}ioinformatics} \textbf{6}(3), \bibinfo{pages}{305--322},
  \dodoi{10.2174/157489311796904637}.

\bibitem[{Lasota(2022)}]{Lasota-DISS2022}
\bibinfo{author}{Lasota, M.} (\textbf{\bibinfo{year}{2022}}).
  \enquote{\bibinfo{title}{Large-eddy simulation for aeroacoustics of human
  phonation}} Ph.D. thesis, \bibinfo{school}{TU Liberec},
  \dodoi{10.13140/RG.2.2.26740.01925}.

\bibitem[{Lasota and {\v{S}}idlof(2019)}]{Lasota-TPFM2019}
\bibinfo{author}{Lasota, M.},  and \bibinfo{author}{{\v{S}}idlof, P.}
  (\textbf{\bibinfo{year}{2019}}). \enquote{\bibinfo{title}{Large-eddy
  simulation of flow through human larynx with a turbulence grid at inlet}} in
  \emph{\bibinfo{booktitle}{Topical Problems of Fluid Mechanics}},
  \dodoi{10.14311/TPFM.2019.020}.

\bibitem[{Lasota \emph{et~al.}(2021)Lasota, {\v{S}}idlof, Kaltenbacher, and
  Schoder}]{Lasota-APPSCI2021}
\bibinfo{author}{Lasota, M.}, \bibinfo{author}{{\v{S}}idlof, P.},
  \bibinfo{author}{Kaltenbacher, M.},  and \bibinfo{author}{Schoder, S.}
  (\textbf{\bibinfo{year}{2021}}). \enquote{\bibinfo{title}{Impact of the
  sub-grid scale turbulence model in aeroacoustic simulation of human voice}}
  \bibinfo{journal}{Applied Sciences} \textbf{11}(4), \bibinfo{pages}{1970},
  \dodoi{10.3390/app11041970}.

\bibitem[{Launchbury(2016)}]{Launchbury-book2016}
\bibinfo{author}{Launchbury, D.~R.} (\textbf{\bibinfo{year}{2016}}).
  \emph{\bibinfo{title}{Unsteady turbulent flow modelling and applications}}
  (\bibinfo{publisher}{Springer: Berlin/Heidelberg}).

\bibitem[{Lesieur \emph{et~al.}(2005)Lesieur, M{\'e}tais, and
  Comte}]{Lesieur-book2005}
\bibinfo{author}{Lesieur, M.}, \bibinfo{author}{M{\'e}tais, O.},  and
  \bibinfo{author}{Comte, P.} (\textbf{\bibinfo{year}{2005}}).
  \emph{\bibinfo{title}{Large-eddy simulations of turbulence}}
  (\bibinfo{publisher}{Cambridge University Press: Cambridge, UK}).

\bibitem[{Lighthill(1952)}]{Lighthill-PRSL1952}
\bibinfo{author}{Lighthill, M.~J.} (\textbf{\bibinfo{year}{1952}}).
  \enquote{\bibinfo{title}{On sound generated aerodynamically. {I}. {General}
  theory}} in \emph{\bibinfo{booktitle}{Proceedings of the {Royal} {Society} of
  {London} {A}: {Mathematical}, {Physical} and {Engineering} {Sciences}}},
  \bibinfo{publisher}{The Royal Society}, Vol. 211, pp.
  \bibinfo{pages}{564--587}, \dodoi{10.1098/rspa.1952.0060}.

\bibitem[{Lodermeyer \emph{et~al.}(2015)Lodermeyer, Becker, D{\"o}llinger, and
  Kniesburges}]{Lodermeyer-EF2015}
\bibinfo{author}{Lodermeyer, A.}, \bibinfo{author}{Becker, S.},
  \bibinfo{author}{D{\"o}llinger, M.},  and \bibinfo{author}{Kniesburges, S.}
  (\textbf{\bibinfo{year}{2015}}). \enquote{\bibinfo{title}{Phase-locked flow
  field analysis in a synthetic human larynx model}}
  \bibinfo{journal}{Experiments in Fluids} \textbf{56}(4),
  \bibinfo{pages}{1--13}, \dodoi{10.1007/s00348-015-1942-6}.

\bibitem[{Mattheus and Br{\"u}cker(2011)}]{Mattheus-JASA2011}
\bibinfo{author}{Mattheus, W.},  and \bibinfo{author}{Br{\"u}cker, C.}
  (\textbf{\bibinfo{year}{2011}}). \enquote{\bibinfo{title}{Asymmetric glottal
  jet deflection: differences of two-and three-dimensional models}}
  \bibinfo{journal}{The Journal of the Acoustical Society of America}
  \textbf{130}(6), \bibinfo{pages}{EL373--EL379}, \dodoi{10.1121/1.3655893}.

\bibitem[{Mihaescu \emph{et~al.}(2010)Mihaescu, Khosla, Murugappan, and
  Gutmark}]{Mihaescu-JASA2010}
\bibinfo{author}{Mihaescu, M.}, \bibinfo{author}{Khosla, S.~M.},
  \bibinfo{author}{Murugappan, S.},  and \bibinfo{author}{Gutmark, E.~J.}
  (\textbf{\bibinfo{year}{2010}}). \enquote{\bibinfo{title}{Unsteady laryngeal
  airflow simulations of the intra-glottal vortical structures}}
  \bibinfo{journal}{Journal of the Acoustical Society of America}
  \textbf{127}(1), \bibinfo{pages}{435--444}, \dodoi{DOI:10.1121/1.3271276}.

\bibitem[{Nicoud and Ducros(1999)}]{Nicoud-FTC1999}
\bibinfo{author}{Nicoud, F.},  and \bibinfo{author}{Ducros, F.}
  (\textbf{\bibinfo{year}{1999}}). \enquote{\bibinfo{title}{Subgrid-scale
  stress modelling based on the square of the velocity gradient tensor}}
  \bibinfo{journal}{Flow, Turbulence and Combustion} \textbf{62}(3),
  \bibinfo{pages}{183--200}, \dodoi{10.1023/A:1009995426001}.

\bibitem[{Pavlidi \emph{et~al.}(2013)Pavlidi, Griffin, Puigt, and
  Mouchtaris}]{Pavlidi-IEEE2013}
\bibinfo{author}{Pavlidi, D.}, \bibinfo{author}{Griffin, A.},
  \bibinfo{author}{Puigt, M.},  and \bibinfo{author}{Mouchtaris, A.}
  (\textbf{\bibinfo{year}{2013}}). \enquote{\bibinfo{title}{Real-time multiple
  sound source localization and counting using a circular microphone array}}
  \bibinfo{journal}{IEEE Transactions on Audio, Speech, and Language
  Processing} \textbf{21}(10), \bibinfo{pages}{2193--2206}.

\bibitem[{Peterson and Barney(1952)}]{Peterson-JASA1952}
\bibinfo{author}{Peterson, G.~E.},  and \bibinfo{author}{Barney, H.~L.}
  (\textbf{\bibinfo{year}{1952}}). \enquote{\bibinfo{title}{Control methods
  used in a study of the vowels}} \bibinfo{journal}{The Journal of the
  Acoustical Society of America} \textbf{24}(2), \bibinfo{pages}{175--184},
  \dodoi{10.1121/1.1906875}.

\bibitem[{Pope(2000)}]{Pope-book2000}
\bibinfo{author}{Pope, S.~B.} (\textbf{\bibinfo{year}{2000}}).
  \emph{\bibinfo{title}{Turbulent Flows}} (\bibinfo{publisher}{Cambridge
  University Press}).

\bibitem[{Rozema \emph{et~al.}(2015)Rozema, Bae, Moin, and
  Verstappen}]{Rozema-PF2015}
\bibinfo{author}{Rozema, W.}, \bibinfo{author}{Bae, H.~J.},
  \bibinfo{author}{Moin, P.},  and \bibinfo{author}{Verstappen, R.}
  (\textbf{\bibinfo{year}{2015}}). \enquote{\bibinfo{title}{Minimum-dissipation
  models for large-eddy simulation}} \bibinfo{journal}{Physics of Fluids}
  \textbf{27}(8), \bibinfo{pages}{085107}, \dodoi{10.1063/1.4928700}.

\bibitem[{Sadeghi \emph{et~al.}(2019{\natexlab{a}})Sadeghi, D{\"o}llinger,
  Kaltenbacher, and Kniesburges}]{Sadeghi-JASA2019}
\bibinfo{author}{Sadeghi, H.}, \bibinfo{author}{D{\"o}llinger, M.},
  \bibinfo{author}{Kaltenbacher, M.},  and \bibinfo{author}{Kniesburges, S.}
  (\textbf{\bibinfo{year}{2019}}{\natexlab{a}}).
  \enquote{\bibinfo{title}{Aerodynamic impact of the ventricular folds in
  computational larynx models}} \bibinfo{journal}{The Journal of the Acoustical
  Society of America} \textbf{145}(4), \bibinfo{pages}{2376--2387},
  \dodoi{10.1121/1.5098775}.

\bibitem[{Sadeghi \emph{et~al.}(2019{\natexlab{b}})Sadeghi, Kniesburges, Falk,
  Kaltenbacher, Sch{\"u}tzenberger, and D{\"o}llinger}]{Sadeghi-APPSCI2019}
\bibinfo{author}{Sadeghi, H.}, \bibinfo{author}{Kniesburges, S.},
  \bibinfo{author}{Falk, S.}, \bibinfo{author}{Kaltenbacher, M.},
  \bibinfo{author}{Sch{\"u}tzenberger, A.},  and
  \bibinfo{author}{D{\"o}llinger, M.}
  (\textbf{\bibinfo{year}{2019}}{\natexlab{b}}).
  \enquote{\bibinfo{title}{Towards a clinically applicable computational larynx
  model}} \bibinfo{journal}{Applied Sciences} \textbf{9}(11),
  \bibinfo{pages}{2288}, \dodoi{10.3390/app9112288}.

\bibitem[{Scherer \emph{et~al.}(2001)Scherer, Shinwari, De~Witt, Zhang,
  Kucinschi, and Afjeh}]{Scherer-JASA2001}
\bibinfo{author}{Scherer, R.}, \bibinfo{author}{Shinwari, D.},
  \bibinfo{author}{De~Witt, J.}, \bibinfo{author}{Zhang, C.},
  \bibinfo{author}{Kucinschi, R.},  and \bibinfo{author}{Afjeh, A.}
  (\textbf{\bibinfo{year}{2001}}). \enquote{\bibinfo{title}{Intraglottal
  pressure profiles for a symmetric and oblique glottis with a divergence angle
  of 10 degrees}} \bibinfo{journal}{The Journal of the Acoustical Society of
  America} \textbf{109}(4), \bibinfo{pages}{1616--1630},
  \dodoi{10.1121/1.1333420}.

\bibitem[{Schoder(2018)}]{Schoder-DISS2018}
\bibinfo{author}{Schoder, S.} (\textbf{\bibinfo{year}{2018}}).
  \enquote{\bibinfo{title}{Aeroacoustic analogies based on compressible flow
  data}} Ph.D. thesis, \bibinfo{school}{TU Wien}.

\bibitem[{Schoder(2022)}]{Schoder-ARXIV2022}
\bibinfo{author}{Schoder, S.} (\textbf{\bibinfo{year}{2022}}).
  \enquote{\bibinfo{title}{{PCWE} for {FSAI}--{D}erivation of scalar wave
  equations for fluid-structure-acoustics interaction of low mach number
  flows}} \dodoi{10.48550/ARXIV.2211.07490}.

\bibitem[{Schoder \emph{et~al.}(2019)Schoder, Junger, Weitz, and
  Kaltenbacher}]{Schoder-AIAA2019}
\bibinfo{author}{Schoder, S.}, \bibinfo{author}{Junger, C.},
  \bibinfo{author}{Weitz, M.},  and \bibinfo{author}{Kaltenbacher, M.}
  (\textbf{\bibinfo{year}{2019}}). \enquote{\bibinfo{title}{Conservative source
  term interpolation for hybrid aeroacoustic computations}} in
  \emph{\bibinfo{booktitle}{25th AIAA/CEAS aeroacoustics conference}}, p.
  \bibinfo{pages}{2538}.

\bibitem[{Schoder and Kaltenbacher(2019)}]{Schoder-JTCA2019}
\bibinfo{author}{Schoder, S.},  and \bibinfo{author}{Kaltenbacher, M.}
  (\textbf{\bibinfo{year}{2019}}). \enquote{\bibinfo{title}{Hybrid aeroacoustic
  computations: State of art and new achievements}} \bibinfo{journal}{Journal
  of Theoretical and Computational Acoustics} \textbf{27}(04),
  \bibinfo{pages}{1950020}, \dodoi{10.1142/s2591728519500208}.

\bibitem[{Schoder \emph{et~al.}(2022)Schoder, Kaltenbacher, Spieser, Vincent,
  Bogey, and Bailly}]{Schoder-AIAA2022}
\bibinfo{author}{Schoder, S.}, \bibinfo{author}{Kaltenbacher, M.},
  \bibinfo{author}{Spieser, {\'E}.}, \bibinfo{author}{Vincent, H.},
  \bibinfo{author}{Bogey, C.},  and \bibinfo{author}{Bailly, C.}
  (\textbf{\bibinfo{year}{2022}}). \enquote{\bibinfo{title}{Aeroacoustic wave
  equation based on {P}ierce's operator applied to the sound generated by a
  mixing layer}} in \emph{\bibinfo{booktitle}{28th AIAA/CEAS Aeroacoustics 2022
  Conference}}, p. \bibinfo{pages}{2896}.

\bibitem[{Schoder \emph{et~al.}(2021{\natexlab{a}})Schoder, Maurerlehner,
  Wurzinger, Hauser, Falk, Kniesburges, D{\"{o}}llinger, and
  Kaltenbacher}]{Schoder-APPSCI2021}
\bibinfo{author}{Schoder, S.}, \bibinfo{author}{Maurerlehner, P.},
  \bibinfo{author}{Wurzinger, A.}, \bibinfo{author}{Hauser, A.},
  \bibinfo{author}{Falk, S.}, \bibinfo{author}{Kniesburges, S.},
  \bibinfo{author}{D{\"{o}}llinger, M.},  and \bibinfo{author}{Kaltenbacher,
  M.} (\textbf{\bibinfo{year}{2021}}{\natexlab{a}}).
  \enquote{\bibinfo{title}{Aeroacoustic sound source characterization of the
  human voice production-perturbed convective wave equation}}
  \bibinfo{journal}{Applied Sciences} \textbf{11}(6),
  \dodoi{10.3390/app11062614}.

\bibitem[{Schoder and Roppert(2022)}]{Schoder-ARXIV2022b}
\bibinfo{author}{Schoder, S.},  and \bibinfo{author}{Roppert, K.}
  (\textbf{\bibinfo{year}{2022}}). \enquote{\bibinfo{title}{open{CFS}: Open
  source finite element software for coupled field simulation--part acoustics}}
  \dodoi{10.48550/ARXIV.2207.04443}.

\bibitem[{Schoder \emph{et~al.}(2020)Schoder, Weitz, Maurerlehner, Hauser,
  Falk, Kniesburges, D{\"o}llinger, and Kaltenbacher}]{Schoder-JASA2020}
\bibinfo{author}{Schoder, S.}, \bibinfo{author}{Weitz, M.},
  \bibinfo{author}{Maurerlehner, P.}, \bibinfo{author}{Hauser, A.},
  \bibinfo{author}{Falk, S.}, \bibinfo{author}{Kniesburges, S.},
  \bibinfo{author}{D{\"o}llinger, M.},  and \bibinfo{author}{Kaltenbacher, M.}
  (\textbf{\bibinfo{year}{2020}}). \enquote{\bibinfo{title}{Hybrid aeroacoustic
  approach for the efficient numerical simulation of human phonation}}
  \bibinfo{journal}{The Journal of the Acoustical Society of America}
  \textbf{147}(2), \bibinfo{pages}{1179--1194}, \dodoi{10.1121/10.0000785}.

\bibitem[{Schoder \emph{et~al.}(2021{\natexlab{b}})Schoder, Wurzinger, Junger,
  Weitz, Freidhager, Roppert, and Kaltenbacher}]{Schoder-JTCA2021}
\bibinfo{author}{Schoder, S.}, \bibinfo{author}{Wurzinger, A.},
  \bibinfo{author}{Junger, C.}, \bibinfo{author}{Weitz, M.},
  \bibinfo{author}{Freidhager, C.}, \bibinfo{author}{Roppert, K.},  and
  \bibinfo{author}{Kaltenbacher, M.}
  (\textbf{\bibinfo{year}{2021}}{\natexlab{b}}).
  \enquote{\bibinfo{title}{Application limits of conservative source
  interpolation methods using a low mach number hybrid aeroacoustic workflow}}
  \bibinfo{journal}{Journal of Theoretical and Computational Acoustics}
  \textbf{29}(01), \bibinfo{pages}{2050032}.

\bibitem[{Schwarze \emph{et~al.}(2011)Schwarze, Mattheus, Klostermann, and
  Br{\"u}cker}]{Schwarze-CF2011}
\bibinfo{author}{Schwarze, R.}, \bibinfo{author}{Mattheus, W.},
  \bibinfo{author}{Klostermann, J.},  and \bibinfo{author}{Br{\"u}cker, C.}
  (\textbf{\bibinfo{year}{2011}}). \enquote{\bibinfo{title}{Starting jet flows
  in a three-dimensional channel with larynx-shaped constriction}}
  \bibinfo{journal}{Computers \& Fluids} \textbf{48}(1),
  \bibinfo{pages}{68--83}, \dodoi{10.1016/j.compfluid.2011.03.016}.

\bibitem[{{\v{S}}idlof \emph{et~al.}(2015){\v{S}}idlof, Z{\"o}rner, and
  H{\"u}ppe}]{Sidlof-BMM2015}
\bibinfo{author}{{\v{S}}idlof, P.}, \bibinfo{author}{Z{\"o}rner, S.},  and
  \bibinfo{author}{H{\"u}ppe, A.} (\textbf{\bibinfo{year}{2015}}).
  \enquote{\bibinfo{title}{A hybrid approach to the computational aeroacoustics
  of human voice production}} \bibinfo{journal}{Biomechanics and Modeling in
  Mechanobiology} \textbf{14}(3), \bibinfo{pages}{473--488},
  \dodoi{10.1007/s10237-014-0617-1}.

\bibitem[{Smagorinsky(1963)}]{Smagorinsky-MWR1963}
\bibinfo{author}{Smagorinsky, J.} (\textbf{\bibinfo{year}{1963}}).
  \enquote{\bibinfo{title}{General circulation experiments with the primitive
  equations: {I}. {T}he basic experiment}} \bibinfo{journal}{Monthly Weather
  Review} \textbf{91}(3), \bibinfo{pages}{99--164},
  \dodoi{10.1175/1520-0493(1963)091<0099:gcewtp>2.3.co;2}.

\bibitem[{Story \emph{et~al.}(1996)Story, Titze, and Hoffman}]{Story-JASA1996}
\bibinfo{author}{Story, B.~H.}, \bibinfo{author}{Titze, I.~R.},  and
  \bibinfo{author}{Hoffman, E.~A.} (\textbf{\bibinfo{year}{1996}}).
  \enquote{\bibinfo{title}{Vocal tract area functions from magnetic resonance
  imaging}} \bibinfo{journal}{The Journal of the Acoustical Society of America}
  \textbf{100}(1), \bibinfo{pages}{537--554}, \dodoi{10.1121/1.415960}.

\bibitem[{Suh and Frankel(2007)}]{Suh-JASA2007}
\bibinfo{author}{Suh, J.},  and \bibinfo{author}{Frankel, S.}
  (\textbf{\bibinfo{year}{2007}}). \enquote{\bibinfo{title}{Numerical
  simulation of turbulence transition and sound radiation for flow through a
  rigid glottal model}} \bibinfo{journal}{Journal of the Acoustical Society of
  America} \textbf{121}(6), \bibinfo{pages}{3728--3739},
  \dodoi{10.1121/1.2723646}.

\bibitem[{Tokuda and Shimamura(2017)}]{Tokuda-JASA2017}
\bibinfo{author}{Tokuda, I.~T.},  and \bibinfo{author}{Shimamura, R.}
  (\textbf{\bibinfo{year}{2017}}). \enquote{\bibinfo{title}{Effect of level
  difference between left and right vocal folds on phonation: Physical
  experiment and theoretical study}} \bibinfo{journal}{The Journal of the
  Acoustical Society of America} \textbf{142}(2), \bibinfo{pages}{482--492},
  \dodoi{10.1121/1.4996105}.

\bibitem[{Versteeg and Malalasekera(2007)}]{Versteeg-book2016}
\bibinfo{author}{Versteeg, H.~K.},  and \bibinfo{author}{Malalasekera, W.}
  (\textbf{\bibinfo{year}{2007}}). \emph{\bibinfo{title}{An introduction to
  computational fluid dynamics: the finite volume method}}
  (\bibinfo{publisher}{Pearson Education: London, UK}).

\bibitem[{Vreugdenhil and Taylor(2018)}]{Vreugdenhil-PF2018}
\bibinfo{author}{Vreugdenhil, C.~A.},  and \bibinfo{author}{Taylor, J.~R.}
  (\textbf{\bibinfo{year}{2018}}). \enquote{\bibinfo{title}{Large-eddy
  simulations of stratified plane couette flow using the anisotropic
  minimum-dissipation model}} \bibinfo{journal}{Physics of Fluids}
  \textbf{30}(8), \bibinfo{pages}{085104}, \dodoi{10.1063/1.5037039}.

\bibitem[{Yoshinaga \emph{et~al.}(2017)Yoshinaga, {Van Hirtum}, and
  Wada}]{Yoshinaga-JSV2017}
\bibinfo{author}{Yoshinaga, T.}, \bibinfo{author}{{Van Hirtum}, A.},  and
  \bibinfo{author}{Wada, S.} (\textbf{\bibinfo{year}{2017}}).
  \enquote{\bibinfo{title}{Multimodal modeling and validation of simplified
  vocal tract acoustics for sibilant /s/}} \bibinfo{journal}{Journal of Sound
  and Vibration} \textbf{411}, \bibinfo{pages}{247--259},
  \dodoi{10.1016/j.jsv.2017.09.004}.

\bibitem[{Yoshizawa and Horiuti(1985)}]{Yoshizawa-PSJ1985}
\bibinfo{author}{Yoshizawa, A.},  and \bibinfo{author}{Horiuti, K.}
  (\textbf{\bibinfo{year}{1985}}). \enquote{\bibinfo{title}{A
  statistically-derived subgrid-scale kinetic energy model for the large-eddy
  simulation of turbulent flows}} \bibinfo{journal}{Journal of the Physical
  Society of Japan} \textbf{54}(8), \bibinfo{pages}{2834--2839},
  \dodoi{10.1143/jpsj.54.2834}.

\bibitem[{Zahiri and Roohi(2019)}]{Zahiri-CF2019}
\bibinfo{author}{Zahiri, A.-P.},  and \bibinfo{author}{Roohi, E.}
  (\textbf{\bibinfo{year}{2019}}). \enquote{\bibinfo{title}{Anisotropic
  minimum-dissipation ({AMD}) subgrid-scale model implemented in openfoam:
  verification and assessment in single-phase and multi-phase flows}}
  \bibinfo{journal}{Computers \& Fluids} \textbf{180},
  \bibinfo{pages}{190--205}, \dodoi{10.1016/j.compfluid.2018.12.011}.

\bibitem[{Zhang \emph{et~al.}(2020)Zhang, Wu, Gray, and Chhetri}]{Zhang-PO2020}
\bibinfo{author}{Zhang, Z.}, \bibinfo{author}{Wu, L.}, \bibinfo{author}{Gray,
  R.},  and \bibinfo{author}{Chhetri, D.~K.} (\textbf{\bibinfo{year}{2020}}).
  \enquote{\bibinfo{title}{Three-dimensional vocal fold structural change due
  to implant insertion in medialization laryngoplasty}} \bibinfo{journal}{PLOS
  ONE} \textbf{15}(1), \bibinfo{pages}{1--11},
  \dodoi{10.1371/journal.pone.0228464}.

\end{thebibliography}



\end{document}